\documentclass[english,12pt]{article}
\usepackage[T1]{fontenc}
\usepackage[latin9]{inputenc}
\usepackage{geometry}
\usepackage{bm}
\usepackage{amsmath}
\usepackage{amsthm}
\usepackage{amssymb}
\usepackage{stmaryrd}
\usepackage{graphicx}
\usepackage{setspace}
\usepackage{color}
\usepackage[authoryear]{natbib}
\makeatletter

\providecommand{\tabularnewline}{\\}

\newcommand{\lyxaddress}[1]{
	\par {\raggedright #1
	\vspace{1.4em}
	\noindent\par}
}
\newenvironment{lyxlist}[1]
	{\begin{list}{}
		{\settowidth{\labelwidth}{#1}
		 \setlength{\leftmargin}{\labelwidth}
		 \addtolength{\leftmargin}{\labelsep}
		 }}
	{\end{list}}
\theoremstyle{plain}
\newtheorem{thm}{\protect\theoremname}
\theoremstyle{plain}
\newtheorem{prop}{\protect\propositionname}
\theoremstyle{plain}
\newtheorem{cor}{\protect\corollaryname}

\usepackage[T1,hyphens]{url}
\usepackage{hyperref}

\@ifundefined{showcaptionsetup}{}{%
 \PassOptionsToPackage{caption=false}{subfig}}
\usepackage{subfig}
\makeatother

\usepackage{babel}
\providecommand{\corollaryname}{Corollary}
\providecommand{\propositionname}{Proposition}
\providecommand{\theoremname}{Theorem}

\begin{document}
\title{Mini-batch learning of exponential family finite mixture models}
\author{Hien D. Nguyen$^{1*}$, Florence Forbes$^{2}$, and Geoffrey J. McLachlan$^{3}$}
\maketitle

\lyxaddress{$^{1}$Department of Mathematics and Statistics, La Trobe University,
Melbourne, Victoria, Australia.\\
$^{2}$Univ. Grenoble Alpes, Inria, CNRS, Grenoble INP$^{\dagger}$,
LJK, 38000 Grenoble, France. $^{\dagger}$Institute of Engineering
Univ. Grenoble Alpes.\\
$^{3}$School of Mathematics and Physics, University of Queensland,
St. Lucia, Brisbane, Australia.\\
$^{*}$Corresponding author: Hien Nguyen (Email: h.nguyen5@latrobe.edu.edu.au).}
\begin{abstract}
Mini-batch algorithms have become increasingly popular due to the
requirement for solving optimization problems, based on large-scale
data sets. Using an existing online expectation-{}-maximization (EM)
algorithm framework, we demonstrate how mini-batch (MB) algorithms
may be constructed, and propose a scheme for the stochastic stabilization
of the constructed mini-batch algorithms. Theoretical results regarding
the convergence of the mini-batch EM algorithms are presented. We
then demonstrate how the mini-batch framework may be applied to conduct
maximum likelihood (ML) estimation of mixtures of exponential family
distributions, with emphasis on ML estimation for mixtures of normal
distributions. Via a simulation study, we demonstrate that the mini-batch
algorithm for mixtures of normal distributions can outperform the
standard EM algorithm. Further evidence of the performance of the
mini-batch framework is provided via an application to the famous
MNIST data set.
\end{abstract}

\section{\label{sec:Introduction}Introduction}

The exponential family of distributions is an important class of probabilistic
models with numerous applications in statistics and machine learning.
The exponential family contains many of the most commonly used univariate
distributions, including the Bernoulli, binomial, gamma, geometric,
inverse Gaussian, logarithmic normal, Poisson, and Rayleigh distributions,
as well as multivariate distributions such as the Dirichlet, multinomial,
multivariate normal, von Mises, and Wishart distributions. See \citet[Ch. 18]{Forbes2011},
\citet[Ch. 18]{DasGupta2011}, and \citet[Ch. 2]{Amari2016}.

Let $\bm{Y}^{\top}=\left(Y_{1},\dots,Y_{d}\right)$ be a random variable
(with realization $\bm{y}$) on the support $\mathbb{Y}\subseteq\mathbb{R}^{d}$
($d\in\mathbb{N}$) , arising from a data generating process (DGP)
with probability density/mass function (PDF/PMF) $f\left(\bm{y};\bm{\theta}\right)$
that is characterized by some parameter vector $\bm{\theta}\in\Theta\subseteq\mathbb{R}^{p}$
($p\in\mathbb{N}$). We say that the distribution that characterizes
the DGP of $\bm{Y}$ is in the exponential family class, if the PDF/PMF
can be written in the form

\begin{equation}
f\left(\bm{y};\bm{\theta}\right)=h\left(\bm{y}\right)\exp\left\{ \left[\bm{s}\left(\bm{y}\right)\right]^{\top}\bm{\phi}\left(\bm{\theta}\right)-\psi\left(\bm{\theta}\right)\right\} \text{,}\label{eq: Exponential family}
\end{equation}
where $\bm{s}\left(\cdot\right)$ and $\bm{\phi}\left(\cdot\right)$
are $p$-dimensional vector functions, and $h\left(\cdot\right)$
and $\psi\left(\cdot\right)$ are $1\text{-dimensional}$ functions
of $\bm{y}$ and $\bm{\theta}$, respectively. If the dimensionality
of $\bm{s}\left(\cdot\right)$ and $\bm{\phi}\left(\cdot\right)$
is less than $p$, then we say that the distribution that characterizes
the DGP of $\bm{Y}$ is in the curved exponential class.

Let $Z\in\left[g\right]$ ($\left[g\right]=\left\{ 1,\dots,g\right\} $;
$g\in\mathbb{N}$) be a latent random variable, and write $\bm{X}^{\top}=\left(\bm{Y}^{\top},Z\right)$.
Suppose that the PDF/PMF of $\left\{ \bm{Y}=\bm{y}|Z=z\right\} $
can be written as $f\left(\bm{y};\bm{\omega}_{z}\right)$, for each
$z\in\left[g\right]$. If we assume that $\mathbb{P}\left(Z=z\right)=\pi_{z}>0$,
such that $\sum_{z=1}^{g}\pi_{z}=1$, then we can write the marginal
PDF/PMF of $\bm{Y}$ in the form

\begin{equation}
f\left(\bm{y};\bm{\theta}\right)=\sum_{z=1}^{g}\pi_{z}f\left(\bm{y};\bm{\omega}_{z}\right)\text{,}\label{eq: Generic mixture}
\end{equation}
where we put the unique elements of $\pi_{z}$ and $\bm{\omega}_{z}$
into $\bm{\theta}$. We call $f\left(\bm{y};\bm{\theta}\right)$ the
$g$-component finite mixture PDF, and we call $f\left(\bm{y};\bm{\omega}_{z}\right)$
the $z\text{th}$ component PDF, characterized by the parameter vector
$\bm{\omega}_{z}\in\Omega$, where $\Omega$ is some subset of a real
product space. We also say that the elements $\pi_{z}$ are prior
probabilities, corresponding to the respective component.

The most common finite mixtures models are mixtures of normal distributions,
which were popularization by \citet{Pearson1894}, and have been prolifically
used by numerous prior authors (cf. \citealp{McLachlan:2019aa}).
The $g$-component $d$-dimensional normal mixture model has PDF of
the form
\begin{equation}
f\left(\bm{y};\bm{\theta}\right)=\sum_{z=1}^{g}\pi_{z}\varphi\left(\bm{y};\bm{\mu}_{z},\bm{\Sigma}_{z}\right)\text{,}\label{eq: normal mixture}
\end{equation}
where the normal PDFs

\begin{equation}
\varphi\left(\bm{y};\bm{\mu}_{z},\bm{\Sigma}_{z}\right)=\left|2\pi\bm{\Sigma}_{z}\right|^{-1/2}\exp\left[-\frac{1}{2}\left(\bm{y}-\bm{\mu}_{z}\right)^{\top}\bm{\Sigma}_{z}^{-1}\left(\bm{y}-\bm{\mu}_{z}\right)\right]\text{,}\label{eq: Normal density}
\end{equation}
replace the component densities $f\left(\bm{y};\bm{\omega}_{z}\right)$,
in (\ref{eq: Generic mixture}). Each component PDF (\ref{eq: Normal density})
is parameterized by a mean vector $\bm{\mu}_{z}\in\mathbb{R}^{d}$
and a positive-definite symmetric  covariance matrix $\bm{\Sigma}_{z}\in\mathbb{R}^{d\times d}$.
We then put each $\pi_{z}$, $\bm{\mu}_{z}$, and $\bm{\Sigma}_{z}$
into the vector $\bm{\theta}$.

As earlier noted, the normal distribution is a member of the exponential
family, and thus (\ref{eq: Normal density}) can be written in form
(\ref{eq: Exponential family}). This can be observed by putting the
unique elements of $\bm{\mu}_{z}$ and $\bm{\Sigma}_{z}$ into $\bm{\omega}_{z}$,
and writing $\varphi\left(\bm{y};\bm{\mu}_{z},\bm{\Sigma}_{z}\right)=f\left(\bm{y};\bm{\omega}_{z}\right)$
in form (\ref{eq: Exponential family}), with mappings

\begin{equation}
h\left(\bm{y}\right)=\left(2\pi\right)^{-d/2}\text{, }\bm{s}\left(\bm{y}\right)=\left[\begin{array}{c}
\bm{y}\\
\textrm{vec}(\bm{y}\bm{y}^{\top})
\end{array}\right]\text{, }\bm{\phi}\left(\bm{\omega}_{z}\right)=\left[\begin{array}{c}
\bm{\Sigma}_{z}^{-1}\bm{\mu}_{z}\\
-\frac{1}{2}\textrm{vec}(\bm{\Sigma}_{z}^{-1})
\end{array}\right]\text{, and}\label{eq: normal exp 1}
\end{equation}
\begin{equation}
\psi\left(\bm{\omega}_{z}\right)=\frac{1}{2}\bm{\mu}_{z}^{\top}\bm{\Sigma}_{z}^{-1}\bm{\mu}_{z}+\frac{1}{2}\log\left|\bm{\Sigma}_{z}\right|\text{.}\label{eq: normal exp 2}
\end{equation}

When conducting data analysis using a normal mixture model, one generally
observes an independent and identically (IID) sequence of $n\in\mathbb{N}$
observations $\left\{ \bm{Y}_{i}\right\} _{i=1}^{n}$, arising from
a DGP that is hypothesized to be characterized by a PDF of the form
(\ref{eq: normal mixture}), with unknown parameter vector $\bm{\theta}=\bm{\theta}_{0}$.
The inferential task is to estimate $\bm{\theta}_{0}$ via some estimator
that is computed from $\left\{ \bm{Y}_{i}\right\} _{i=1}^{n}$. The
most common computational approach to obtaining an estimator of $\bm{\theta}_{0}$
is via maximum likelihood (ML) estimation, using the expectation-{}-maximization
algorithm (EM; \citealp{Dempster1977}). See \citet[Ch. 3.2]{McLachlan2000}
for a description of the normal mixture EM algorithm. Generally, when
$g$, $d$, and $n$ are of small to moderate size, the conventional
EM approach is feasible, and is able to perform the task of ML estimation
in a timely manner. Unfortunately, due to its high memory demands,
costly matrix operations \citep{NguyenMcLachlan2015}, and slow convergence
rates \citep[Sec. 3.9]{McLachlan2008}, the conventional EM algorithm
is not suited for the computational demands of analyzing increasingly
large data sets, such as those that could be considered as \emph{big
data} in volumes such as \citet{Buhlmann2016}, \citet{Han2017},
and \citet{Hardle2018}.

Over the years, numerous algorithms have been proposed as means to
alleviate the computational demands of the EM algorithm for normal
mixture models. Some of such approaches include the component-wise
algorithm of \citet{Celeux2001}, the greedy algorithm of \citet{Vlassis2002},
the sparse and incremental $kd$-tree algorithm of \citet{Ng2004},
the subspace projection algorithm of \citet{Bouveyron2007}, and the
matrix operations-free algorithm of \citet{NguyenMcLachlan2015}.

There has been a recent resurgence in stochastic approximation algorithms,
of the \citet{Robbins1951} and \citet{Kiefer1952} type, developed
for the purpose of solving computationally challenging optimization
problems, such as the ML estimation of normal mixture models. A good
review of the current literature can be found in \citet{Chau2015}.
Naïve and direct applications of the stochastic approximation approach
to mixture model estimation can be found in \citet{Liang2008}, \citet{Zhang2008},
and \citet{Nguyen:2018aa}. 

Following a remark from \citet{Cappe2009} regarding the possible
extensions of the online EM algorithm, we propose mini-batch EM algorithms
for the ML estimation of exponential family mixture models. These
algorithms include a number of variants, among which are update truncation
variants that had not been made explicit, before. Using the theorems
from \citet{Cappe2009}, we state results regarding the convergence
of our algorithms. We then specialize our attention to the important
case of normal mixture models, and demonstrate that the required assumptions
for convergence are met in such a scenario.

A thorough numerical study is conducted in order to assess the performance
of our normal mixture mini-batch algorithms. Comparisons are drawn
between our algorithms and the usual batch EM algorithm for ML estimation
of normal mixture models. We show that our mini-batch algorithms can
be applied to very large data sets by demonstrating its applicability
to the ML estimation of normal mixture models on the famous MNIST
data of \citet{LeCun1998}.

References regarding mixtures of exponential family distributions
and EM-type stochastic approximation algorithms, and comments regarding
some recent related literature are relegated to the Supplementary
Materials, in the interest of brevity. Additional remarks, numerical
results, and derivations are also included in these Supplementary
Materials in order to provide extra context and further demonstrate
the capabilities of the described framework. These demonstrations
include the derivation of mini-batch EM algorithms for mixtures of
exponential and Poisson distributions. The Supplementary Materials
can be found at \url{https://github.com/hiendn/StoEMMIX/blob/master/Manuscript_files/SupplementaryMaterials.pdf}.

The remainder of the paper is organized as follows. In Section 2,
we present the general results of \citet{Cappe2009} and demonstrate
how they can be used for mini-batch ML estimation of exponential family
mixture models. In Section 3, we derive the mini-batch EM algorithms
for the ML estimation of normal mixtures, as well as verify the convergence
of the algorithms using the results of \citet{Cappe2009}. Via numerical
simulations, we compare the performance of our mini-batch algorithms
to the usual EM algorithm for ML estimation of normal mixture models,
in Section 4. A set of real data study on a very large data set is
presented in Section 5. Conclusions are drawn in Section 6. Additional
material, such as mini-batch EM algorithms for exponential and Poisson
mixture models, can be found in the Supplementary Materials.

\section{\label{sec:The-minibatch-EM}The mini-batch EM algorithm}

Suppose that we observe a single pair of random variables $\bm{X}^{\top}=\left(\bm{Y}^{\top},\bm{Z}^{\top}\right)$,
where $\bm{Y}\in\mathbb{Y}$ is observed but $\bm{Z}\in\mathbb{L}$
is latent, where $\mathbb{Y}$ and $\mathbb{L}$ are subsets of multivariate
real-valued spaces. Furthermore, suppose that the marginal PDF/PMF
of $\bm{Y}$ is hypothesized to be of the form $f\left(\bm{y};\bm{\theta}_{0}\right)$,
for some unknown parameter vector $\bm{\theta}_{0}\in\Theta\subseteq\mathbb{R}^{p}$.
A good estimator for $\bm{\theta}_{0}$ is the ML estimator $\hat{\bm{\theta}}$
that can be defined as:

\begin{equation}
\hat{\bm{\theta}}\in\left\{ \hat{\bm{\theta}}:\log f\left(\bm{Y};\hat{\bm{\theta}}\right)=\max_{\bm{\theta}\in\Theta}\log f\left(\bm{Y};\bm{\theta}\right)\right\} \text{.}\label{eq: ML definition}
\end{equation}

When the problem (\ref{eq: ML definition}) cannot be solved in a
simple manner (e.g. when the solution does not exist in closed form),
one may seek to employ an iterative scheme in order to obtain an ML
estimator. If the joint PDF/PMF of $\bm{X}$ is known, then one can
often construct an EM algorithm in order to solve the problem in the
bracket of (\ref{eq: ML definition}).

Start with some initial guess for $\bm{\theta}_{0}$ and call it the
zeroth iterate of the EM algorithm $\bm{\theta}^{\left(0\right)}$
and suppose that we can write the point PDF/PMF of $\bm{X}$ as $f\left(\bm{y},\bm{z};\bm{\theta}\right)$,
for any $\bm{\theta}$. At the $r\text{th}$ iterate of the EM algorithm,
we perform an expectation (E-) step, followed by a maximization (M-)
step. The $r\text{th}$ E-step consists of obtaining the conditional
expectation of the complete-data log-likelihood (i.e. $\log f\left(\bm{y},\bm{z};\bm{\theta}\right)$)
given the observed data, using the current estimate of the parameter
vector

\[
Q\left(\bm{\theta};\bm{\theta}^{\left(r-1\right)}\right)=\mathbb{E}_{\bm{\theta}^{\left(r-1\right)}}\left[\log f\left(\bm{y},\bm{Z};\bm{\theta}\right)|\bm{Y}=\bm{y}\right]\text{,}
\]
which we will call the conditional expected complete-data log-likelihood.

Upon obtaining the conditional expectation of the complete-data log-likelihood,
one then conducts the $r\text{th}$ M-step by solving the problem
\[
\bm{\theta}^{\left(r\right)}=\arg\underset{\bm{\theta}\in\Theta}{\max}\text{ }Q\left(\bm{\theta};\bm{\theta}^{\left(r-1\right)}\right)\text{.}
\]
The E- and M-steps are repeated until some stopping criterion is met.
Upon termination, the final iterate of the algorithm is taken as a
solution for problem (\ref{eq: ML definition}). See \citet{McLachlan2008}
for a thorough exposition regarding the EM algorithm.

\subsection{\label{subsec:The-online-EM}The online EM algorithm}

Suppose that we observe a sequence of $n$ IID replicates of the variable
$\bm{Y}$, $\left\{ \bm{Y}_{i}\right\} _{i=1}^{n}$, where each $\bm{Y}_{i}$
is the visible component of the pair $\bm{X}_{i}=\left(\bm{Y}_{i}^{\top},\bm{Z}_{i}^{\top}\right)$
($i\in\left[n\right]$). In the online learning context, each of the
observations from $\left\{ \bm{Y}_{i}\right\} _{i=1}^{n}$ is observed
one at a time, in sequential order.

Using the sequentially obtained sequence $\left\{ \bm{Y}_{i}\right\} _{i=1}^{n}$,
we wish to obtain an ML estimator for the parameter vector $\bm{\theta}_{0}$,
in the same sense as in (\ref{eq: ML definition}). In order to construct
an online EM algorithm framework with provable convergence, \citet{Cappe2009}
assume the following restrictions regarding the nature of the hypothesized
DGP of $\left\{ \bm{Y}_{i}\right\} _{i=1}^{n}$.
\begin{lyxlist}{00.00.0000}
\item [{A1}] The complete-data likelihood corresponding to the pair $\bm{X}$
is of exponential family form. That is,
\begin{equation}
f\left(\bm{x};\bm{\theta}\right)=h\left(\bm{x}\right)\exp\left\{ \left[\bm{s}\left(\bm{x}\right)\right]^{\top}\bm{\phi}\left(\bm{\theta}\right)-\psi\left(\bm{\theta}\right)\right\} \text{,}\label{eq: Complete data EF}
\end{equation}
where $h\left(\cdot\right)$, $\psi\left(\cdot\right)$, $\bm{s}\left(\cdot\right)$,
and $\bm{\phi}\left(\cdot\right)$ are as defined for (\ref{eq: Exponential family}).
\item [{A2}] The function
\begin{equation}
\bar{\bm{s}}\left(\bm{y};\bm{\theta}\right)=\mathbb{E}_{\bm{\theta}}\left[\bm{s}\left(\bm{X}\right)|\bm{Y}=\bm{y}\right]\label{eq: sbar}
\end{equation}
is well defined for all $\bm{y}\in\mathbb{Y}$ and $\bm{\theta}\in\Theta$.
\item [{A3}] There exists a convex open subset $\mathbb{S}\subseteq\mathbb{R}^{p}$,
which satisfies the properties that:
\begin{lyxlist}{00.00.0000}
\item [{(i)}] for all $\bm{s}\in\mathbb{S}$, $\bm{y}\in\mathbb{Y}$, $\bm{\theta}\in\Theta$,
$\left(1-\gamma\right)\bm{s}+\gamma\bar{\bm{s}}\left(\bm{y};\bm{\theta}\right)\in\mathbb{S}$
for any $\gamma\in\left(0,1\right)$, and
\item [{(ii)}] for any $\bm{s}\in\mathbb{S}$, the function
\[
q\left(\bm{s};\bm{\theta}\right)=\bm{s}^{\top}\bm{\phi}\left(\bm{\theta}\right)-\psi\left(\bm{\theta}\right)
\]
has a unique global maximum over $\Theta$, which will be denoted
by
\[
\bar{\bm{\theta}}\left(\bm{s}\right)=\arg\underset{\bm{\theta}\in\Theta}{\max}\text{ }q\left(\bm{s};\bm{\theta}\right)\text{.}
\]
\end{lyxlist}
\end{lyxlist}
Let $Q_{n}\left(\bm{\theta};\bm{\theta}^{\left(r-1\right)}\right)$
be the expected complete-data log-likelihood over data $\left\{ \bm{Y}_{i}\right\} _{i=1}^{n}$,
at the $r\text{th}$ E-step of an EM algorithm for solving the problem:
\[
\hat{\bm{\theta}}_{n}\in\left\{ \hat{\bm{\theta}}:n^{-1}\sum_{i=1}^{n}\log f\left(\bm{Y}_{i};\hat{\bm{\theta}}\right)=\max_{\bm{\theta}\in\Theta}n^{-1}\sum_{i=1}^{n}\log f\left(\bm{Y}_{i};\bm{\theta}\right)\right\} \text{,}
\]
where we say that $\hat{\bm{\theta}}_{n}$ is the ML estimator, based
on the data $\left\{ \bm{Y}_{i}\right\} _{i=1}^{n}$. When, A1--A3
are satisfied, we can write
\[
Q_{n}\left(\bm{\theta};\bm{\theta}^{\left(r-1\right)}\right)=nq\left(n^{-1}\sum_{i=1}^{n}\bar{\bm{s}}\left(\bm{Y}_{i};\bm{\theta}^{\left(r-1\right)}\right);\bm{\theta}\right)+\text{Constant}\text{,}
\]
which can then be maximized, with respect to $\bm{\theta}$, in order
to yield an M-step update of the form:
\begin{equation}
\bm{\theta}^{\left(r\right)}=\bar{\bm{\theta}}\left(n^{-1}\sum_{i=1}^{n}\bar{\bm{s}}\left(\bm{Y}_{i};\bm{\theta}^{\left(r-1\right)}\right)\right)\text{,}\label{eq: theta_bar generic}
\end{equation}
where $\bm{\theta}^{\left(r\right)}$ is a function that depends only
on the average $n^{-1}\sum_{i=1}^{n}\bar{\bm{s}}\left(\bm{Y}_{i};\bm{\theta}^{\left(r-1\right)}\right)$.

Now we suppose that we sample the individual observations of $\left\{ \bm{Y}_{i}\right\} _{i=1}^{n}$,
one at a time and sequentially. Furthermore, upon observation of $\bm{Y}_{i}$,
we wish to compute an online estimate of $\bm{\theta}_{0}$, which
we  denote as $\bm{\theta}^{\left(i\right)}$. Based on  the simplification
of the EM algorithm under A1--A3, as described above, \citet{Cappe2009}
proposed the following online EM algorithm.

Upon observation of $\bm{Y}_{i}$, compute the intermediate updated
sufficient statistic
\begin{equation}
\bm{s}^{\left(i\right)}=\bm{s}^{\left(i-1\right)}+\gamma_{i}\left[\bar{\bm{s}}\left(\bm{Y}_{i};\bm{\theta}^{\left(i-1\right)}\right)-\bm{s}^{\left(i-1\right)}\right]\text{,}\label{eq: Update Sufficient}
\end{equation}
with $\bm{s}^{\left(0\right)}=\bar{\bm{s}}\left(\bm{Y}_{i};\bm{\theta}^{\left(0\right)}\right)$.
Here, $\gamma_{i}$ is the $i\text{th}$ term of the learning rate
sequence that we will discuss in further details in the sequel. Observe
that we can also write
\[
\bm{s}^{\left(i\right)}=\gamma_{i}\bar{\bm{s}}\left(\bm{Y}_{i};\bm{\theta}^{\left(i-1\right)}\right)+\left(1-\gamma_{i}\right)\bm{s}^{\left(i-1\right)}\text{,}
\]
which makes it clear that for $\gamma_{i}\in\left(0,1\right)$, $\bm{s}^{\left(i\right)}$
is a weighted average between $\bar{\bm{s}}\left(\bm{Y}_{i};\bm{\theta}^{\left(i-1\right)}\right)$
and $\bm{s}^{\left(i-1\right)}$. Using $\bm{s}^{\left(i\right)}$
and the function $\bar{\bm{\theta}}$, we can then express the $i\text{th}$
iteration online EM estimate of $\bm{\theta}_{0}$ as
\begin{equation}
\bm{\theta}^{\left(i\right)}=\bar{\bm{\theta}}\left(\bm{s}^{\left(i\right)}\right)\text{.}\label{eq: Generic EM update}
\end{equation}

Next, we state a consistency theorem that strongly motivates the use
of the online EM algorithm, defined by (\ref{eq: Update Sufficient})
and (\ref{eq: Generic EM update}). Suppose that the true DGP that
generates each $\bm{Y}_{i}$ of $\left\{ \bm{Y}_{i}\right\} _{i=1}^{n}$
is characterized by the probability measure $F_{0}$. Write the expectation
operator with respect to this measure as $\mathbb{E}_{F_{0}}$. In
order to state the consistency result of \citet{Cappe2009}, we require
the following additional set of assumptions.
\begin{lyxlist}{00.00.0000}
\item [{B1}] The parameter space $\Theta$ is a convex and open subset
of a real product space, and the functions $\bm{\phi}$ and $\psi$,
in (\ref{eq: Complete data EF}), are both twice continuously differentiable
with respect to $\bm{\theta}\in\Theta$.
\item [{B2}] The function $\bar{\bm{\theta}}$, as defined in (\ref{eq: theta_bar generic}),
is a continuously differentiable function with respect to $\bm{s}\in\mathbb{S}$,
where $\mathbb{S}$ is as defined in A3.
\item [{B3}] For some $p>2$, and all compact $\mathbb{K}\subset\mathbb{S}$,
\begin{equation}
\sup_{\bm{s}\in\mathbb{K}}\text{ }\mathbb{E}_{F_{0}}\left[\left|\bar{\bm{s}}\left(\bm{Y};\bar{\bm{\theta}}\left(\bm{s}\right)\right)\right|^{p}\right]<\infty\text{.}\label{eq: B3 eq}
\end{equation}
\end{lyxlist}
As the algorithm defined by (\ref{eq: Update Sufficient}) and (\ref{eq: Generic EM update})
is of the Robbins-Monro type, establishment of convergence of the
algorithm requires the definition of a mean field (see \citealp{Chen:2003aa}
and \citealp{Kushner:2003aa} for comprehensive treatments regarding
such algorithms). In the case of the online EM algorithm, we write
the mean field as
\[
\bm{h}\left(\bm{s}\right)=\mathbb{E}_{F_{0}}\left[\bar{\bm{s}}\left(\bm{Y};\bar{\bm{\theta}}\left(\bm{s}\right)\right)\right]-\bm{s}
\]
and define the set of its roots as $\Gamma=\left\{ \bm{s}\in\mathbb{S}:\bm{h}\left(\bm{s}\right)=\bm{0}\right\} $.

Define the log-likelihood of the hypothesized PDF $f\left(\cdot;\bm{\theta}\right)$
with respect to the measure $F_{0}$, as
\[
\ell\left(f\left(\cdot;\bm{\theta}\right)\right)=\mathbb{E}_{F_{0}}\left[\log f\left(\bm{Y};\bm{\theta}\right)\right]\text{.}
\]

Let $\nabla_{\bm{\theta}}$ denote the gradient with respect to $\bm{\theta}$,
and define the sets
\[
\mathbb{W}_{\Gamma}=\left\{ \ell\left(f\left(\cdot;\bm{\theta}\right)\right):\bm{\theta}=\bar{\bm{\theta}}\left(\bm{s}\right)\text{, }\bm{s}\in\Gamma\right\} 
\]
and
\[
\mathbb{M}_{\Theta}=\left\{ \hat{\bm{\theta}}\in\Theta:\nabla_{\bm{\theta}}\left.\ell\left(f\left(\cdot;\bm{\theta}\right)\right)\right|_{\bm{\theta}=\hat{\bm{\theta}}}=\bm{0}\right\} \text{.}
\]

Note that $\mathbb{M}_{\Theta}$ is the set of stationary points of
the log-likelihood function. Further, define the distance between
a real vector $\bm{a}$ and a set $\mathbb{B}$ by
\[
\textrm{dist}\left(\bm{a},\mathbb{B}\right)=\inf_{\bm{b}\in\mathbb{B}}\left\Vert \bm{a}-\bm{b}\right\Vert \text{,}
\]
where $\left\Vert \cdot\right\Vert $ is the usual Euclidean metric,
and denote the complement of a subset $\mathbb{A}$ of a real product
space by $\mathbb{A}^{c}$. Finally, make the following assumptions.
\begin{lyxlist}{00.00.0000}
\item [{C1}] The sequence of learning rates $\left\{ \gamma_{i}\right\} _{i=1}^{\infty}$
fulfills the conditions that $0<\gamma_i<1$, for each $i$,
\[
\sum_{i=1}^{\infty}\gamma_{i}=\infty\text{, and }\sum_{i=1}^{\infty}\gamma_{i}^{2}<\infty\text{.}
\]
\item [{C2}] At initialization $\bm{s}^{\left(0\right)}\in\mathbb{S}$
and, with probability 1,
\[
\limsup_{i\rightarrow\infty}\left|\bm{s}^{\left(i\right)}\right|<\infty\text{, and }\liminf_{i\rightarrow\infty}\text{ }\textrm{dist}\left(\bm{s}^{\left(i\right)},\mathbb{S}^{c}\right)=0\text{.}
\]
\item [{C3}] The set $\mathbb{W}_{\Gamma}$ is nowhere dense.
\end{lyxlist}
\begin{thm}
[Cappe and Moulines, 2009] \label{thm cappe moulines} Assume that
A1--A3, B1--B3, and C1--C3 are satisfied, and let $\left\{ \bm{Y}_{i}\right\} _{i=1}^{\infty}$
be an IID sample with DGP characterized by the PDF $f_{0}$, which
is hypothesized to have the form $f\left(\cdot;\bm{\theta}\right)$,
as in (\ref{eq: Complete data EF}). Further, let $\left\{ \bm{s}^{\left(i\right)}\right\} _{i=1}^{\infty}$
and $\left\{ \bm{\theta}^{\left(i\right)}\right\} _{i=1}^{\infty}$
be sequences generated by the online EM algorithm, defined by (\ref{eq: Update Sufficient})
and (\ref{eq: Generic EM update}). Then, with probability 1,
\[
\lim_{i\rightarrow\infty}\text{ dist}\left(\bm{s}^{\left(i\right)},\Gamma\right) = 0\text{, and }\lim_{i\rightarrow\infty}\text{ }\textrm{dist}\left(\bm{\theta}^{\left(i\right)},\mathbb{M}_{\Theta}\right)= 0\text{.}
\]
\end{thm}
Notice that this result allows for a mismatch between the true probability
measure $F_{0}$ and the assumed pseudo-true family $f\left(\cdot;\bm{\theta}\right)$
from which $\left\{ \bm{Y}_{i}\right\} _{i=1}^{\infty}$ is hypothesized
to arise. This therefore allows for misspecification, in the sense
of \citet{White1982}, which is almost certain to occur in the modeling
of any sufficiently complex data. In any case, the online EM algorithm
will converge towards an estimate of the parameter vector $\bm{\theta}$,
which is in the set $\mathbb{M}_{\Theta}$. When the DGP can be characterized
by a density in the family of the form $f\left(\cdot;\bm{\theta}\right)$,
we observe that $\mathbb{M}_{\Theta}$ contains not only the global
maximizer of the log-likelihood function, but also local maximizers,
minimizers, and saddle points. Thus, the online algorithm suffers
from the same lack of strong convergence guarantees, as the batch
EM algorithm (cf. \citealp{Wu1983}).

In the case of misspecification the set $\mathbb{M}_{\Theta}$ will
include the parameter vector $\bm{\theta}_{0}$ that maximizes the
log-likelihood function, with respect to the true probability measure
$F_{0}$. However, as with the well-specified case, it will also include
stationary points of other types, as well. We further provide characterizations
of the sets $\mathbb{W}_{\Gamma}$ and $\mathbb{M}_{\Theta}$ in terms
of the Kullback-Leibler divergence (KL; \citealp{KullbackLeibler1951})
in the Supplementary Materials.

Assumption C1 can be fulfilled by taking sequences $\left\{ \gamma_{i}\right\} _{i=1}^{\infty}$
of form $\gamma_{i}=\gamma_{0}i^{\alpha}$, for some $\alpha\in\left(0,1\right]$
and $\gamma_{0}\in\left(0,1\right)$. We shall discuss this point
further, in the sequel. Although the majority of the assumptions can
be verified or are fulfilled by construction, the two limits in C2
stand out as being particularly difficult to verify. In \citet{Cappe2009},
the authors suggest that one method for enforcing C2 is to use the
method of update truncation, but they did not provide an explicit
scheme for conducting such truncation.

A truncation version of the algorithm defined by (\ref{eq: Update Sufficient})
and (\ref{eq: Generic EM update}) can be specified via the method
of \citet{Delyon1999}. That is, let $\left\{ \mathbb{K}_{m}\right\} _{m=0}^{\infty}$
be a sequence of compact sets, such that

\begin{equation}
\mathbb{K}_{m}\subset\text{interior}\left(\mathbb{K}_{m+1}\right)\text{, and }\bigcup_{m=0}^{\infty}\mathbb{K}_{m}=\mathbb{S}\text{.}\label{eq: Km seq}
\end{equation}
We then replace (\ref{eq: Update Sufficient}) and (\ref{eq: Generic EM update})
by the following scheme. At the $i\text{th}$ iteration, firstly compute
\begin{equation}
\tilde{\bm{s}}^{\left(i\right)}=\bm{s}^{\left(i-1\right)}+\gamma_{i}\left[\bar{\bm{s}}\left(\bm{Y}_{i};\bm{\theta}^{\left(i-1\right)}\right)-\bm{s}^{\left(i-1\right)}\right]\text{.}\label{eq: generic trunc1}
\end{equation}
Secondly,
\begin{equation}
\text{if }\tilde{\bm{s}}^{\left(i\right)}\in\mathbb{K}_{m_{i-1}}\text{, then set }\bm{s}^{\left(i\right)}=\tilde{\bm{s}}^{\left(i\right)}\text{, }\bm{\theta}^{\left(i\right)}=\bar{\bm{\theta}}\left(\bm{s}^{\left(i\right)}\right)\text{, and }m_{i}=m_{i-1}\text{,}\label{eq: generic trunc2}
\end{equation}
else
\begin{equation}
\text{if }\tilde{\bm{s}}^{\left(i\right)}\notin\mathbb{K}_{m_{i-1}}\text{, then set }\bm{s}^{\left(i\right)}=\bm{S}_{i}\text{, }\bm{\theta}^{\left(i\right)}=\bar{\bm{\theta}}\left(\bm{S}_{i}\right)\text{, and }m_{i}=m_{i-1}+1\text{,}\label{eq: generic trunc3}
\end{equation}
where $\left\{ \bm{S}_{i}\right\} _{i=1}^{\infty}$ is an arbitrary
random sequence, such that $\bm{S}_{i}\in\mathbb{K}_{0}$, for each
$i\in\mathbb{N}$. We have the following result regarding the algorithm
defined by (\ref{eq: generic trunc1})--(\ref{eq: generic trunc3}).
\begin{prop}
\label{prop: Truncated EM}Assume that A1--A3, B1--B3, C1 and C3
are satisfied, and let $\left\{ \bm{Y}_{i}\right\} _{i=1}^{\infty}$
be an IID sample with DGP characterized by the PDF $f_{0}$, which
is hypothesized to have the form $f\left(\cdot;\bm{\theta}\right)$,
as in (\ref{eq: Complete data EF}). Further, let $\left\{ \bm{s}^{\left(i\right)}\right\} _{i=1}^{\infty}$
and $\left\{ \bm{\theta}^{\left(i\right)}\right\} _{i=1}^{\infty}$
be sequences generated by the truncated online EM algorithm, defined
by (\ref{eq: generic trunc1})--(\ref{eq: generic trunc3}). Then,
with probability 1,
\[
\lim_{i\rightarrow\infty}\text{ dist}\left(\bm{s}^{\left(i\right)},\Gamma\right)=0\text{, and }\lim_{i\rightarrow\infty}\text{ }\textrm{dist}\left(\bm{\theta}^{\left(i\right)},\mathbb{M}_{\Theta}\right)=0\text{.}
\]
\end{prop}
The proof of Proposition \ref{prop: Truncated EM} requires the establishment
of equivalence between A1--A3, B1--B3, C1, and C3, and the many
assumptions of Theorem 3 and 6 of \citet{Delyon1999}. Thus the proof
is simple and mechanical, but long and tedious. We omit it for the
sake of brevity.

\subsection{The mini-batch algorithm}

At the most elementary level, a mini-batch algorithm for computation
of a sequence of estimators $\left\{ \bm{\theta}^{\left(r\right)}\right\} _{r=1}^{R}$
for some parameter $\bm{\theta}_{0}$, from some sample $\left\{ \bm{Y}_{i}\right\} _{i=1}^{n}$,
where $R\in\mathbb{N}$, has the following property. The
algorithm is iterative, and at the $r\text{th}$ iteration of the
algorithm, the estimator $\bm{\theta}^{\left(r\right)}$ only depends
on the previous iterate $\bm{\theta}^{\left(r-1\right)}$ and some
subsample, possibly with replacement, of $\left\{ \bm{Y}_{i}\right\} _{i=1}^{n}$.
Typical examples of mini-batch algorithms include the many variants
of the stochastic gradient descent-class of algorithms; see, for example,
\citet{Cotter:2011aa}, \citet{Li:2014aa}, \citet{Zhao:2014aa},
and \citet{Ghadimi:2016aa}.

Suppose that we observe a fixed size realization $\left\{ \bm{y}_{i}\right\} _{i=1}^{n}$
of some IID random sample $\left\{ \bm{Y}\right\} _{i=1}^{n}$. Furthermore,
fix a so-called batch size $N\le n$ and a learning rate sequence
$\left\{ \gamma_{r}\right\} _{r=1}^{R}$, and select some appropriate
initial values $\bm{s}^{\left(0\right)}$ and $\bm{\theta}^{\left(0\right)}$
from which the sequences $\left\{ \bm{s}^{\left(r\right)}\right\} _{r=1}^{R}$
and $\left\{ \bm{\theta}^{\left(r\right)}\right\} _{r=1}^{R}$ can
be constructed. A mini-batch version of the online EM algorithm, specified
by (\ref{eq: Update Sufficient}) and (\ref{eq: Generic EM update})
can be specified as follows. For each $r\in\left[R\right]$, sample
$N$ observations from $\left\{ \bm{y}_{i}\right\} _{i=1}^{n}$ uniformly,
with replacement, and denote the subsample by $\left\{ \bm{Y}_{i}^{r}\right\} _{i=1}^{N}$.
Then, using $\left\{ \bm{Y}_{i}^{r}\right\} _{i=1}^{N}$, compute

\begin{equation}
\bm{s}^{\left(r\right)}=\bm{s}^{\left(r-1\right)}+\gamma_{r}\left[N^{-1}\sum_{i=1}^{N}\bar{\bm{s}}\left(\bm{Y}_{i}^{r};\bm{\theta}^{\left(r-1\right)}\right)-\bm{s}^{\left(r-1\right)}\right]\text{, and }\bm{\theta}^{\left(r\right)}=\bar{\bm{\theta}}\left(\bm{s}^{\left(r\right)}\right)\text{.}\label{eq: MB EM generic}
\end{equation}

In order to justify the mini-batch algorithm, we make the following
observation. The online EM algorithm, defined by (\ref{eq: Update Sufficient})
and (\ref{eq: Generic EM update}), is designed to obtain a root in
the set $\mathbb{M}_{\Theta}$, which is a vector $\hat{\bm{\theta}}\in\Theta$
such that
\[
\nabla_{\bm{\theta}}\left.\ell\left(f\left(\cdot;\bm{\theta}\right)\right)\right|_{\bm{\theta}=\hat{\bm{\theta}}}=\bm{0}\text{.}
\]
If $N=1$ (i.e., the case proposed in \citealp[Sec. 2.5]{Cappe2009}),
then the DGP for generating subsamples is simply a single draw from
the empirical measure:
\[
F_{\text{Emp}}\left(\bm{y}\right)=\sum_{i=1}^{n}\frac{1}{n}\delta\left(\bm{y}-\bm{y}_{i}\right)\text{,}
\]
where $\delta$ is the Dirac delta function (see, for details, \citealp[Ch. 2]{Prosperetti2011}).
We can write
\begin{align}
\ell\left(f\left(\cdot;\bm{\theta}\right)\right) & =\mathbb{E}_{F_{0}}\left[\log f\left(\bm{Y};\bm{\theta}\right)\right]\nonumber \\
 & =\mathbb{E}_{F_{\text{Emp}}}\left[\log f\left(\bm{Y};\bm{\theta}\right)\right]\nonumber \\
 & =\frac{1}{n}\sum_{i=1}^{n}\log f\left(\bm{y}_{i};\bm{\theta}\right)\text{,}\label{eq: KL Emp 1}
\end{align}
which is the log-likelihood function, with respect to the realization
$\left\{ \bm{y}_{i}\right\} _{i=1}^{n}$, under the density function
of form $f\left(\cdot;\bm{\theta}\right)$. Thus, in the $N=1$ case,
the algorithm defined by (\ref{eq: MB EM generic}) solves for  log-likelihood
roots $\hat{\bm{\theta}}$ of the form
\[
\frac{1}{n}\sum_{i=1}^{n}\nabla_{\bm{\theta}}\left.\log f\left(\bm{y}_{i};\bm{\theta}\right)\right|_{\bm{\theta}=\hat{\bm{\theta}}}=\mathbf{0},
\]
or equivalently, solving for an element in the set
\[
\mathbb{M}_{\Theta}^{\text{Emp}}=\left\{ \hat{\bm{\theta}}\in\Theta:\sum_{i=1}^{n}\nabla_{\bm{\theta}}\left.\log f\left(\bm{y}_{i};\bm{\theta}\right)\right|_{\bm{\theta}=\hat{\bm{\theta}}}=\mathbf{0}\right\} \text{.}
\]
The $N>1$ case follows the same argument, and is described in the
Supplementary Materials (Section 2.2). 
Let $F_{\text{Emp}}^{N}$ denote the probability measure corresponding
to the DGP of $N$ independent random samples from $F_{\text{Emp}}$.
We have the following result, based
on Theorem \ref{thm cappe moulines}.
\begin{cor}
\label{cor minibatch generic}For any $N\in\mathbb{N}$, assume that
A1--A3, B1--B3, and C1--C3 are satisfied (replacing $i$ by $r$,
and $F_{0}$ by $F_{\text{Emp}}^{N}$, where appropriate), and let
$\left\{ \bm{y}_{i}\right\} _{i=1}^{n}$ be a realization of some
IID random sequence $\left\{ \bm{Y}_{i}\right\} _{i=1}^{n}$, where
each $\bm{Y}_{i}$ is hypothesized to arise from a DGP having PDF
of the form $f\left(\cdot;\bm{\theta}\right)$, as in (\ref{eq: Complete data EF}).
Let $\left\{ \bm{s}^{\left(r\right)}\right\} _{i=1}^{\infty}$ and
$\left\{ \bm{\theta}^{\left(r\right)}\right\} _{i=1}^{\infty}$ be
sequences generated by the mini-batch EM algorithm, defined by (\ref{eq: Update Sufficient})
and (\ref{eq: Generic EM update}). Then, with probability 1,
\[
\lim_{r\rightarrow\infty}\text{ dist}\left(\bm{s}^{\left(r\right)},\Gamma\right)=0\text{, and }\lim_{r\rightarrow\infty}\text{ }\textrm{dist}\left(\bm{\theta}^{\left(r\right)},\mathbb{M}_{\Theta}^{\text{Emp}}\right)=0\text{.}
\]
\end{cor}
That is, as we take $R\rightarrow\infty$, the algorithm defined by
(\ref{eq: Update Sufficient}) and (\ref{eq: Generic EM update})
will identify elements in the sets $\Gamma$ and $\mathbb{M}_{\Theta}^{\text{Emp}}$,
with probability 1. As with the case of Theorem \ref{thm cappe moulines},
C2 is again difficult to verify. Let $\left\{ \mathbb{K}_{m}\right\} _{m=0}^{\infty}$
be as per (\ref{eq: Km seq}). Then, we replace the algorithm defined
via (\ref{eq: MB EM generic}), by the following truncated version.

Again, suppose that we observe a fixed size realization $\left\{ \bm{y}_{i}\right\} _{i=1}^{n}$
of some IID random sample $\left\{ \bm{Y}\right\} _{i=1}^{n}$. Furthermore,
fix a so-called batch size $N\le n$ and a learning rate sequence
$\left\{ \gamma_{r}\right\} _{r=1}^{R}$, and select some appropriate
initial values $\bm{s}^{\left(0\right)}$ and $\bm{\theta}^{\left(0\right)}$
from which the sequences $\left\{ \bm{s}^{\left(r\right)}\right\} _{r=1}^{R}$
and $\left\{ \bm{\theta}^{\left(r\right)}\right\} _{r=1}^{R}$ can
be constructed. For each $r\in\left[R\right]$, sample $N$ observations
from $\left\{ \bm{y}_{i}\right\} _{i=1}^{n}$ uniformly, with replacement,
and denote the subsample by $\left\{ \bm{Y}_{i}^{r}\right\} _{i=1}^{N}$.
Using $\left\{ \bm{Y}_{i}^{r}\right\} _{i=1}^{N}$, compute

\begin{equation}
\tilde{\bm{s}}^{\left(r\right)}=\bm{s}^{\left(r-1\right)}+\gamma_{r}\left[N^{-1}\sum_{i=1}^{N}\bar{\bm{s}}\left(\bm{Y}_{i}^{r};\bm{\theta}^{\left(r-1\right)}\right)-\bm{s}^{\left(r-1\right)}\right]\text{.}\label{eq: minibatch trunc gen}
\end{equation}
Then, with $i$ being appropriately replaced by $r$, use (\ref{eq: generic trunc2})
and (\ref{eq: generic trunc3}) to compute $\bm{s}^{\left(r\right)}$
and $\bm{\theta}^{\left(r\right)}$. We obtain the following result
via an application of Proposition \ref{prop: Truncated EM}.
\begin{cor}
\label{cor trunc}For any $N\in\mathbb{N}$, assume that A1--A3,
B1--B3, and C1--C3 are satisfied (replacing $i$ by $r$, and $F_{0}$
by $F_{\text{Emp}}^{N}$, where appropriate), and let $\left\{ \bm{y}_{i}\right\} _{i=1}^{n}$
be a realization of some IID random sequence $\left\{ \bm{Y}_{i}\right\} _{i=1}^{n}$,
where each $\bm{Y}_{i}$ is hypothesized to arise from a DGP having
PDF of the form $f\left(\cdot;\bm{\theta}\right)$, as in (\ref{eq: Complete data EF}).
Let $\left\{ \bm{s}^{\left(r\right)}\right\} _{i=1}^{\infty}$ and
$\left\{ \bm{\theta}^{\left(r\right)}\right\} _{i=1}^{\infty}$ be
sequences generated by the truncated mini-batch EM algorithm, defined
by (\ref{eq: minibatch trunc gen}), (\ref{eq: generic trunc2}),
and (\ref{eq: generic trunc3}). Then, with probability 1,
\[
\lim_{r\rightarrow\infty}\text{ dist}\left(\bm{s}^{\left(r\right)},\Gamma\right)=0\text{, and }\lim_{r\rightarrow\infty}\text{ }\textrm{dist}\left(\bm{\theta}^{\left(r\right)},\mathbb{M}_{\Theta}^{\text{Emp}}\right)=0\text{.}
\]
\end{cor}

\subsection{The learning rate sequence}

As previously stated, a good choice for the learning rate sequence
$\left\{ \gamma_{i}\right\} _{i=1}^{\infty}$ is to take $\gamma_{i}=\gamma_{0}i^{\alpha}$,
for each $i\in\mathbb{N}$, such that $\alpha\in\left(1/2,1\right]$
and $\gamma_{0}\in\left(0,1\right)$. Under the assumptions of Theorem
\ref{thm cappe moulines}, \citet[Thm. 2]{Cappe2009} showed that
the learning rate choice leads to the convergence of the sequence
$\gamma_{0}^{1/2}i^{\alpha/2}\left(\bm{\theta}^{\left(i\right)}-\bm{\theta}_{0}\right)$,
in distribution, to a normal distribution with mean $\bm{0}$ and
covariance matrix depending on $\bm{\theta}_{0}$, for some $\bm{\theta}_{0}\in\mathbb{M}_{\Theta}$.
Here $\left\{ \bm{\theta}^{\left(i\right)}\right\} _{i=1}^{\infty}$
is a sequence of online EM algorithm iterates, generated by (\ref{eq: Update Sufficient})
and (\ref{eq: Generic EM update}). A similar result can be stated
for the truncated online EM, mini-batch EM, and truncated mini-batch
EM algorithms, by replacing the relevant indices and quantities in
the previous statements by their respective counterparts.

The result above implies that the convergence rate is $\bm{\theta}^{\left(i\right)}-\bm{\theta}_{0}=o_{\text{p}}\left(i^{\alpha/2}\right)$,
for any valid $\alpha$, where $o_{\text{p}}$ is the usual order
in probability notation (see \citealp[Defn. 2.33]{White2001}). Thus,
it would be tempting to take $\alpha=1$ in order to obtain a rate
with optimal order of $n^{1/2}$. However, as shown in \citet[Thm. 2]{Cappe2009},
the $\alpha=1$ case requires constraints on $\gamma_{0}$ in order
to fulfill a stability assumption that is impossible to validate,
in practice.

It is, however, still possible to obtain a sequence of estimators
that converges to some $\bm{\theta}_{0}$ at a rate with optimal order
$n^{1/2}$. We can do this via the famous so-called Polyak averaging
scheme of \citet{Polyak:1990aa} and \citet{Polyak:1992aa}. In the
current context, one takes as an input the sequence of online EM iterates
$\left\{ \bm{\theta}^{\left(i\right)}\right\} _{i=1}^{\infty}$, and
output the running average sequence $\left\{ \bm{\theta}_{\text{A}}^{\left(i\right)}\right\} _{i=1}^{\infty}$,
where
\begin{equation}
\bm{\theta}_{\text{A}}^{\left(i\right)}=i^{-1}\sum_{j=1}^{i}\bm{\theta}^{\left(j\right)}\text{,}\label{eq: Polyak}
\end{equation}
for each $i\in\mathbb{N}$. For any $\alpha\in\left(1/2,1\right)$,
it is provable that $\bm{\theta}_{\text{A}}^{\left(i\right)}-\bm{\theta}_{0}=o_{\text{p}}\left(n^{1/2}\right)$.
As before, this result generalizes to the cases of the truncated online
EM, mini-batch EM, and truncated mini-batch EM algorithms, also.

We note that the computation of the $i\text{th}$ running average
term (\ref{eq: Polyak}) does not require the storage of the entire
sequence of iterates $\left\{ \bm{\theta}^{\left(i\right)}\right\} _{i=1}^{\infty}$,
as one would anticipate by applying (\ref{eq: Polyak}) naïvely. One
can instead write (\ref{eq: Polyak}) in the iterative form
\[
\bm{\theta}_{\text{A}}^{\left(i\right)}=i^{-1}\left[\left(i-1\right)\bm{\theta}_{\text{A}}^{\left(i-1\right)}+\bm{\theta}^{\left(i\right)}\right]\text{.}
\]

\section{\label{sec:Normal-mixture-models}Normal mixture models}

\subsection{Finite mixtures of exponential family distributions}

We recall from Section \ref{sec:Introduction}, that the random variable
$\bm{Y}$ is said to arise from a DGP characterized by a $g$ component
finite mixture of component PDFs of form $f\left(\bm{y};\bm{\omega}_{z}\right)$,
if it has a PDF of the form (\ref{eq: Generic mixture}). Furthermore,
if the component PDFs are of the exponential family form (\ref{eq: Exponential family}),
then we  further write the PDF of $\bm{Y}$ as

\begin{equation}
f\left(\bm{y};\bm{\theta}\right)=\sum_{z=1}^{g}\pi_{z}h\left(\bm{y}\right)\exp\left\{ \left[\bm{s}\left(\bm{y}\right)\right]^{\top}\bm{\phi}\left(\bm{\omega}_{z}\right)-\psi\left(\bm{\omega}_{z}\right)\right\} \text{.}\label{eq: finite mixture expfam}
\end{equation}

From the construction of the finite mixture model, we have the fact
that (\ref{eq: finite mixture expfam}) is the marginalization of
the joint density of the random variable $\bm{X}^{\top}=\left(\bm{Y}^{\top},Z\right)$:

\begin{equation}
f\left(\bm{x};\bm{\theta}\right)=\prod_{\zeta=1}^{g}\left[\pi_{\zeta}h\left(\bm{y}\right)\exp\left\{ \left[\bm{s}\left(\bm{y}\right)\right]^{\top}\bm{\phi}\left(\bm{\omega}_{\zeta}\right)-\psi\left(\bm{\omega}_{\zeta}\right)\right\} \right]^{\left\llbracket z=\zeta\right\rrbracket }\label{eq: complete data gen}
\end{equation}
over the random variable $Z\in\left[g\right]$, recalling that $Z$
is a categorical random variable with $g$ categories (cf. \citealp[Ch. 2]{McLachlan2000}).
Here, $\left\llbracket c\right\rrbracket $ is the Iverson bracket
notation, that takes value 1 if condition $c$ is true, and 0 otherwise
\citep[Ch. 1]{Iverson:1967aa}. We  rewrite (\ref{eq: complete data gen})
as follows:

\begin{align*}
f\left(\bm{x};\bm{\theta}\right) & =h\left(\bm{y}\right)\exp\left\{ \sum_{\zeta=1}^{g}\left\llbracket z=\zeta\right\rrbracket \left[\log\pi_{\zeta}+\left[\bm{s}\left(\bm{y}\right)\right]^{\top}\bm{\phi}\left(\bm{\omega}_{\zeta}\right)-\psi\left(\bm{\omega}_{\zeta}\right)\right]\right\} \\
 & =h\left(\bm{x}\right)\exp\left\{ \left[\bm{s}\left(\bm{x}\right)\right]^{\top}\bm{\phi}\left(\bm{\theta}\right)-\psi\left(\bm{\theta}\right)\right\} \text{,}
\end{align*}
where $h\left(\bm{x}\right)=h\left(\bm{y}\right)$, $\psi\left(\bm{\theta}\right)=0$,

\[
\bm{s}\left(\bm{x}\right)=\left[\begin{array}{c}
\left\llbracket z=1\right\rrbracket \\
\left\llbracket z=1\right\rrbracket \bm{s}\left(\bm{y}\right)\\
\vdots\\
\left\llbracket z=g\right\rrbracket \\
\left\llbracket z=g\right\rrbracket \bm{s}\left(\bm{y}\right)
\end{array}\right]\text{, and }\bm{\phi}\left(\bm{\theta}\right)=\left[\begin{array}{c}
\log\pi_{1}-\psi\left(\bm{\omega}_{1}\right)\\
\bm{\phi}\left(\bm{\omega}_{1}\right)\\
\vdots\\
\log\pi_{g}-\psi\left(\bm{\omega}_{g}\right)\\
\bm{\phi}\left(\bm{\omega}_{g}\right)
\end{array}\right]\text{,}
\]
and thus obtain the following general result regarding finite mixtures
of exponential family distributions.
\begin{prop}
\label{prop complete data rep}The complete-data likelihood of any
finite mixture of exponential family distributions with PDF of the
form (\ref{eq: finite mixture expfam}) can also be written in the
exponential family form (\ref{eq: Complete data EF}).
\end{prop}
With Proposition \ref{prop complete data rep}, we have proved that
when applying the online EM or the mini-batch EM algorithm to the
problem of conducting ML estimation for any finite mixture model of
exponential family distributions, A1 is automatically satisfied.

\subsection{Finite mixtures of normal distributions}

Recall from Section \ref{sec:Introduction} that the random variable
$\bm{Y}$ is said to be distributed according to a $g\text{-component}$
finite mixture of normal distributions, if it characterized by a PDF
of the form (\ref{eq: normal mixture}). Using the exponential family
decomposition from (\ref{eq: normal exp 1}) and (\ref{eq: normal exp 2}),
we  write the complete-data likelihood of $\bm{X}^{\top}=\left(\bm{Y}^{\top},Z\right)$
in the form (\ref{eq: Complete data EF}) by setting $h\left(\bm{x}\right)=\left(2\pi\right)^{-d/2}$,
$\psi\left(\bm{\theta}\right)=0$,
\begin{equation}
\bm{s}\left(\bm{x}\right)=\left[\begin{array}{c}
\left\llbracket z=1\right\rrbracket \\
\left\llbracket z=1\right\rrbracket \bm{y}\\
\left\llbracket z=1\right\rrbracket \textrm{vec}( \bm{y}\bm{y}^{\top})\\
\vdots\\
\left\llbracket z=g\right\rrbracket \\
\left\llbracket z=g\right\rrbracket \bm{y}\\
\left\llbracket z=g\right\rrbracket \textrm{vec}(\bm{y}\bm{y}^{\top})
\end{array}\right]\text{, and }\bm{\phi}\left(\bm{\theta}\right)=\left[\begin{array}{c}
\log\pi_{1}-\frac{1}{2}\bm{\mu}_{1}^{\top}\bm{\Sigma}_{1}^{-1}\bm{\mu}_{1}+\frac{1}{2}\log\left|\bm{\Sigma}_{1}\right|\\
\bm{\Sigma}_{1}^{-1}\bm{\mu}_{1}\\
-\frac{1}{2} \textrm{vec}(\bm{\Sigma}_{1}^{-1})\\
\vdots\\
\log\pi_{g}-\frac{1}{2}\bm{\mu}_{g}^{\top}\bm{\Sigma}_{g}^{-1}\bm{\mu}_{g}+\frac{1}{2}\log\left|\bm{\Sigma}_{g}\right|\\
\bm{\Sigma}_{g}^{-1}\bm{\mu}_{g}\\
-\frac{1}{2}\textrm{vec}(\bm{\Sigma}_{g}^{-1})
\end{array}\right]\text{,}\label{eq: s and psi norm}
\end{equation}
where $\textrm{vec}(\cdot)$ is the matrix vectorization operator.

Using the results from \citet[Ch. 3]{McLachlan2000}, we  write the
conditional expectation (\ref{eq: sbar}) in the form
\[
\left[\bar{\bm{s}}\left(\bm{y};\bm{\theta}\right)\right]^{\top}=\left(\tau_{1}\left(\bm{y};\bm{\theta}\right),\tau_{1}\left(\bm{y};\bm{\theta}\right)\bm{y},\tau_{1}\left(\bm{y};\bm{\theta}\right) \textrm{vec}(\bm{y}\bm{y}^{\top}),\dots,\tau_{g}\left(\bm{y};\bm{\theta}\right),\tau_{g}\left(\bm{y};\bm{\theta}\right)\bm{y},\tau_{g}\left(\bm{y};\bm{\theta}\right)\textrm{vec}(\bm{y}\bm{y}^{\top}\right))\text{,}
\]
where
\[
\tau_{z}\left(\bm{y};\bm{\theta}\right)=\frac{\pi_{z}\varphi\left(\bm{y};\bm{\mu}_{z},\bm{\Sigma}_{z}\right)}{\sum_{\zeta=1}^{g}\pi_{\zeta}\varphi\left(\bm{y};\bm{\mu}_{\zeta},\bm{\Sigma}_{\zeta}\right)}\text{,}
\]
is the usual \emph{a posteriori} probability that $Z=z$ ($z\in\left[g\right]$),
given observation of $\bm{Y}=\bm{y}$. Again, via the results from
\citet[Ch. 3]{McLachlan2000}, we  write the update function $\bar{\bm{\theta}}$
in the following form. Define $\bar{\bm{\theta}}$ to have the elements
$\bar{\pi}_{z}$ and $\bar{\bm{\omega}}_{z}$, for each $z\in\left[g\right]$,
where each $\bar{\bm{\omega}}_{z}$ subsequently has elements $\bar{\bm{\mu}}_{z}$
and $\bar{\bm{\Sigma}}_{z}$. Furthermore, for convenience, we define for $\bm{s}$ the following notation
\[
\bm{s}^\top = (s_{11}, \bm{s}_{21}, \textrm{vec}(\mathbf{S}_{31}),\dots, s_{1g}, \bm{s}_{2g}, \textrm{vec}(\mathbf{S}_{3g}))\text{,}
\]
and 
\[
\left[\bar{\bm{s}}\left(\bm{y};\bm{\theta}\right)\right]^\top= (\bar{s}_{11}\left(\bm{y};\bm{\theta}\right), \bar{\bm{s}}_{21}\left(\bm{y};\bm{\theta}\right), \textrm{vec}(\bar{\mathbf{S}}_{31}\left(\bm{y};\bm{\theta}\right)),\dots,\bar{s}_{1g}\left(\bm{y};\bm{\theta}\right), \bar{\bm{s}}_{2g}\left(\bm{y};\bm{\theta}\right), \textrm{vec}(\bar{\mathbf{S}}_{3g}\left(\bm{y};\bm{\theta}\right)))\text{,}
\]
with 
\[
\bar{s}_{1z}\left(\bm{y};\bm{\theta}\right)=\tau_{z}\left(\bm{y};\bm{\theta}\right)\text{, }\bar{\bm{s}}_{2z}\left(\bm{y};\bm{\theta}\right)=\tau_{z}\left(\bm{y};\bm{\theta}\right)\bm{y}\text{, and }\bar{\mathbf{S}}_{3z}\left(\bm{y};\bm{\theta}\right)=\tau_{z}\left(\bm{y};\bm{\theta}\right)\bm{y}\bm{y}^{\top}\text{.}
\]
Then the application of the M-step is equivalent to apply function  $\bar{\bm{\theta}}$ as a function of $\bm{s}$, countaining the unique elements of $\bar{\pi}_{z}$,  $\bar{\bm{\mu}}_{z}$, and $\bar{\bm{\Sigma}}_{z}$, for $z\in[g]$, 
defined by
\begin{equation}
\bar{\pi}_{z}(\bm{s})=\frac{s_{1z}}{\sum_{j=1}^g s_{1j}}, \quad 
\bar{\bm{\mu}}_{z}(\bm{s}) = \frac{ \bm{s}_{2z}}{ s_{1z}}, \quad \mbox{and }
 \bar{\bm{\Sigma}}_{z}(\bm{s}) = \frac{\mathbf{S}_{3z}}{s_{1z}}-\frac{\bm{s}_{2z} \bm{s}_{2z}^{\top}}{s_{1z}^{2}}\text{.}
 \label{eq:thetabargauss}
\end{equation}



This implies that the mini-batch EM and truncated mini-batch
EM algorithms proceed via update rule $\bar{\bm{\theta}}\left(\bm{s}^{\left(r\right)}\right)$, where $\bar{\bm{\theta}}$ and $\bm{s}^{\left(r\right)}$ are as given in 
(\ref{eq: MB EM generic}). We start from $\bm{\theta}^{(0)}$ and 
$\bm{s}^{(1)}=N^{-1} \sum_{i=1}^N \bar{\bm{s}}(\bm{Y}_i, \bm{\theta}^{(0)}) $. 
Then,
$\bm{\theta}^{(1)\top}= [\bar{\bm{\theta}}(\bm{s}^{(1)})]^\top\text{,}$
which has elements
\begin{equation}
\bar{\pi}_{z}\left(\bm{s}^{\left(1\right)}\right)=N^{-1}\sum_{i=1}^{N}\tau_{z}\left(\bm{Y}_{i};\bm{\theta}^{\left(0\right)}\right)\text{, } \quad 
\bar{\bm{\mu}}_{z}\left(\bm{s}^{\left(1\right)}\right)=\frac{\sum_{i=1}^{N}{\tau}_{z}\left(\bm{Y}_{i};\bm{\theta}^{\left(0\right)}\right) \bm{Y}_{i}}{\sum_{i=1}^{N}\tau_{z}\left(\bm{Y}_{i};\bm{\theta}^{\left(0\right)}\right)}\text{,}\label{eq: N batch updates1}
\end{equation}
and
\begin{equation}
\bar{\bm{\Sigma}}_{z}\left(\bm{s}^{\left(1\right)}\right)=\frac{\sum_{i=1}^{N}\mathbf{\tau}_{z}\left(\bm{Y}_{i};\bm{\theta}^{\left(0\right)}\right)\bm{Y}_{i}\bm{Y}_{i}^{\top} }{\sum_{i=1}^{N}\tau_{z}\left(\bm{Y}_{i};\bm{\theta}^{\left(0\right)}\right)}-\frac{\left[\sum_{i=1}^{N}\tau_{z}\left(\bm{Y}_{i};\bm{\theta}^{\left(0\right)}\right) \bm{Y}_{i}\right]\left[\sum_{i=1}^{N}\tau_{z}\left(\bm{Y}_{i};\bm{\theta}^{\left(0\right)}\right) \bm{Y}_{i}\right]^{\top}}{\left[\sum_{i=1}^{N}\tau_{z}\left(\bm{Y}_{i};\bm{\theta}^{\left(0\right)}\right)\right]^{2}}\text{.}\label{eq: N batch updates2}
\end{equation}

\subsection{Convergence analysis of the mini-batch algorithm}

In addition to Assumptions A1--A3, B1--B3, and C1--C3, make the
additional assumption
\begin{lyxlist}{00.00.0000}
\item [{D1}] The Hessian matrix of $\sum_{i=1}^{n}\log f\left(\bm{y}_{i};\bm{\theta}\right)$,
evaluated at any $\bm{\theta}_{0}\in\mathbb{M}_{\Theta}^{\text{Emp}}$,
is non-singular with respect to $\bm{\theta}\in\Theta$.
\end{lyxlist}
Assumption D1 is generally satisfied for all but pathological samples
$\left\{ \bm{y}_{i}\right\} _{i=1}^{n}$. The following result is
proved in the Supplementary Materials.
\begin{prop}
\label{prop norm}Let $\left\{ \bm{y}_{i}\right\} _{i=1}^{n}$ be
a realization of some IID random sequence $\left\{ \bm{Y}_{i}\right\} _{i=1}^{n}$,
where each $\bm{Y}_{i}$ is hypothesized to arise from a DGP having
PDF of the form (\ref{eq: normal mixture}). If $\left\{ \bm{s}^{\left(r\right)}\right\} _{i=1}^{\infty}$
and $\left\{ \bm{\theta}^{\left(r\right)}\right\} _{i=1}^{\infty}$
are sequences generated by the mini-batch EM algorithm, defined by
(\ref{eq: MB EM generic}) and (\ref{eq:thetabargauss}), 
 then for any $N\in\mathbb{N}$, if
C1, C2, and D1 are satisfied (replacing $i$ by $r$, and $F_{0}$
by $\prod_{j=1}^{N}F_{\text{Emp}}$, where appropriate), then, with
probability 1,
\[
\lim_{r\rightarrow\infty}\text{ dist}\left(\bm{s}^{\left(r\right)},\Gamma\right)=0\text{, and }\lim_{r\rightarrow\infty}\text{ }\textrm{dist}\left(\bm{\theta}^{\left(r\right)},\mathbb{M}_{\Theta}^{\text{Emp}}\right)=0\text{.}
\]
Alternatively, if $\left\{ \bm{s}^{\left(r\right)}\right\} _{i=1}^{\infty}$
and $\left\{ \bm{\theta}^{\left(r\right)}\right\} _{i=1}^{\infty}$
are sequences generated by the truncated mini-batch EM algorithm,
defined by (\ref{eq: minibatch trunc gen}), (\ref{eq: generic trunc2}),
(\ref{eq: generic trunc3}) and (\ref{eq:thetabargauss}),
then for any $N\in\mathbb{N}$, if C1 and D1 are satisfied (replacing
$i$ by $r$, and $F_{0}$ by $\prod_{j=1}^{N}F_{\text{Emp}}$, where
appropriate), then, with probability 1,
\[
\lim_{r\rightarrow\infty}\text{ dist}\left(\bm{s}^{\left(r\right)},\Gamma\right)=0\text{, and }\lim_{r\rightarrow\infty}\text{ }\textrm{dist}\left(\bm{\theta}^{\left(r\right)},\mathbb{M}_{\Theta}^{\text{Emp}}\right)=0\text{.}
\]
\end{prop}

\subsection{A truncation sequence}

In order to apply the truncated version of the mini-batch EM algorithm,
we require an appropriate sequence $\left\{ \mathbb{K}_{m}\right\} _{m=0}^{\infty}$
that satisfies condition (\ref{eq: Km seq}). This can be constructed
in parts. Let us write

\begin{equation}
\mathbb{K}_{m}=\mathbb{D}_{g-1}^{m}\times\prod_{i=1}^{g}\left(\mathbb{B}_{d}^{m}\times \mathbb{H}_{d}^{m}\right)\text{,}\label{eq: K for normal}
\end{equation}
where we shall let $c_{1},c_{2},c_{3}\ge1$,
\[
\mathbb{D}_{g-1}^{m}=\left\{ \left(\pi_{1},\dots,\pi_{g}\right)\in\mathbb{R}^{g}:\sum_{z=1}^{g}\pi_{z}=1\text{, and }\pi_{z}\ge\frac{1}{c_{1}+m}\text{, for each }z\in\left[g\right]\right\} \text{,}
\]
\[
\mathbb{B}_{d}^m=\left[-\left(c_{2}+m\right),c_{2}+m\right]^d\text{,}
\]
and
\[
\mathbb{H}_{d}^{m}=\left\{ \mathbf{H}\in\mathbb{H}_{d}:\lambda_{1}\left(\mathbf{H}\right)\ge\frac{1}{c_{3}+m}\text{, }\lambda_{d}\left(\mathbf{H}\right)\le c_{3}+m\right\} \text{,}
\]
using the notation $\lambda_{1}\left(\mathbf{H}\right)$ and $\lambda_{d}\left(\mathbf{H}\right)$
to denote the smallest and largest eigenvalues of the matrix $\mathbf{H}$.
A justification regarding this truncation scheme can be found in the
Supplementary Materials.

We make a final note that the construction (\ref{eq: K for normal})
is not a unique method for satisfying the conditions of (\ref{eq: Km seq}).
One can instead, for example, replace $c_{j}+m$, by $c_{j}\left(1+m\right)$
($j\in\left[3\right]$) in the definitions of the sets that constitute
(\ref{eq: K for normal}).

\section{\label{sec:Simulation-studies}Simulation studies}

We present a pair of simulation studies, based upon the famous Iris
data set of \citet{Fisher1936} and the Wreath data of \citet{Fraley:2005aa},
in the main text. A further four simulation scenarios are presented
in the Supplementary Materials. In each case, we utilize the initial
small data sets, obtained from the base $\texttt{R}$ package \citep{RCT2018}
and the $\texttt{mclust}$ package for $\texttt{R}$ \citep{Scrucca:2016aa},
respectively, and use them as templates to generate much larger data
sets. All computations are conducted in the $\texttt{R}$ programming
environment, although much of the bespoke programs are programmed
in $\texttt{C}$ and integrated in $\texttt{R}$ via the $\texttt{Rcpp}$
and $\texttt{RcppArmadillo}$ packages of \citep{Eddelbuettel2013}.
Furthermore, timings of programs were conducted on a MacBook Pro with
a 2.2 GHz Intel Core i7 processor, 16 GB of 1600 MHz DDR3 RAM, and
a 500 GB SSD hard drive. We note that all of the code used to conduct
the simulations and computations for this manuscript can be accessed
from \url{https://github.com/hiendn/StoEMMIX}.

In the sequel, in all instances, we shall use the learning rate sequence
$\left\{ \gamma_{r}\right\} _{r=1}^{\infty}$, where $\gamma_{r}=\left(1-10^{-10}\right)\times r^{6/10}$,
which follows from the choice made by \citet{Cappe2009} in their
experiments. In all computations, a fixed number of epochs (or epoch
equivalence) of $10$ is allotted to each algorithm. Here, recall
that the number of epochs is equal to the number of sweeps through
the data set $\left\{ \bm{y}_{i}\right\} _{i=1}^{n}$ that an algorithm
is allowed. Thus, drawing $10n$ observations from the data $\left\{ \bm{y}_{i}\right\} _{i=1}^{n}$,
with replacement, is equivalent to 10 epochs. Thus, each iteration
of the standard EM algorithm counts as a single epoch, whereas, for
a mini-batch algorithm with batch size $N$, every $n/N$ iterations
counts as an epoch.

Next, in both of our studies, we consider batch sizes of $N=n/10$
and $N=n/5$, we further consider Polyak averaging as well as truncation.
Thus, for each study, a total of eight variants of the mini-batch
EM algorithm is considered. In the truncate case, we set $c_{1},c_{2},c_{3}=1000$.
Finally, the variants of the mini-batch EM algorithm are compared
to the standard (batch) EM algorithm for fitting finite mixtures of
normal distributions. In the interest of fairness, each of the algorithms
is initialized at the same starting value of $\bm{\theta}^{\left(0\right)}$,
using the randomized initialization scheme suggested in \citet[Sec. 3.9.3]{McLachlan2000}.
That is, the same randomized starting instance is used for the EM
algorithm and each of the mini-batch variants.

To the best of our knowledge, the most efficient and reliable implementation
of the EM algorithm for finite mixtures of normal distributions, in
$\texttt{R}$, is the $\text{\texttt{em}}$ function from the $\texttt{mclust}$
package. Thus, this will be used for all of our comparisons. Data
generation from the template data sets is handled using the $\texttt{simdataset}$
function from the $\texttt{MixSim}$ package \citep{Melnykov:2012aa},
in the Iris study, and the $\texttt{simVVV}$ function from $\texttt{mclust}$
in the Wreath study. Timing was conducted using the $\mathtt{proc.time}$
function.

\subsection{Iris data}

The Iris data (accessed in $\texttt{R}$ via the $\texttt{data(iris)}$
command) contain measurements of $d=4$ dimensions from 150 iris
flowers, 50 of each are of the species \emph{Setosa}, \emph{Versicolor},
and \emph{Virginica}, respectively. The 4 dimensions of each flower
that are measured are \emph{petal length}, \emph{petal width}, \emph{sepal
length}, and \emph{sepal width}. To each of the subpopulations of
species, we fit a single multivariate normal distribution to the 50
observations (i.e., we estimate a mean vector and covariance matrix,
for each species). Then, using the three mean vectors and covariance
matrices, we construct a template $g=3$ component normal mixture
model with equal mixing proportions $\pi_{z}=1/3$ ($z\in\left[3\right]$),
of form (\ref{eq: normal mixture}). This template distribution is
then used to generate synthetic data sets of any size $n$.

Two experiments are performed using this simulation scheme. In the
first experiment, we generate $n=10^{6}$ observations $\left\{ \bm{y}_{i}\right\} _{i=1}^{n}$
from the template. We then utilize $\left\{ \bm{y}_{i}\right\} _{i=1}^{n}$
and each of the truncated EM algorithm variants as well as the batch
EM algorithm to compute ML estimates. We use a number of measures
of performance for each algorithm variant. These include the computation
time, the log-likelihood, the squared error of the parameter estimates
(SE; the Euclidean distance as compared to the generative parameter
vector), and the adjusted-Rand index (ARI; \citealp{Hubert1985})
between the maximum \emph{a posteriori} clustering labels obtained
from the fitted mixture model and the true generative data labels.

The ARI measures whether or not two sets of labels are in concordance
or not. Here a value of 1 indicates perfect similarity, and 0 indicates
discordance. Since the ARI allows for randomness in the labelling
process, it is possible to have negative ARI values, which are rare
and also indicates discordance in the data. Each variant is repeated
$\text{Rep}=100$ times, and each performance measurement is recorded
in order to obtain a measure of the overall performance of each algorithm.
For future reference, we name this study Iris1. In the second study,
which we name Iris2, we repeat the setup of Iris1 but with the number
of observations increased to $n=10^{7}$.

\subsection{Wreath data}

\begin{figure}
\begin{centering}
\includegraphics[width=15cm]{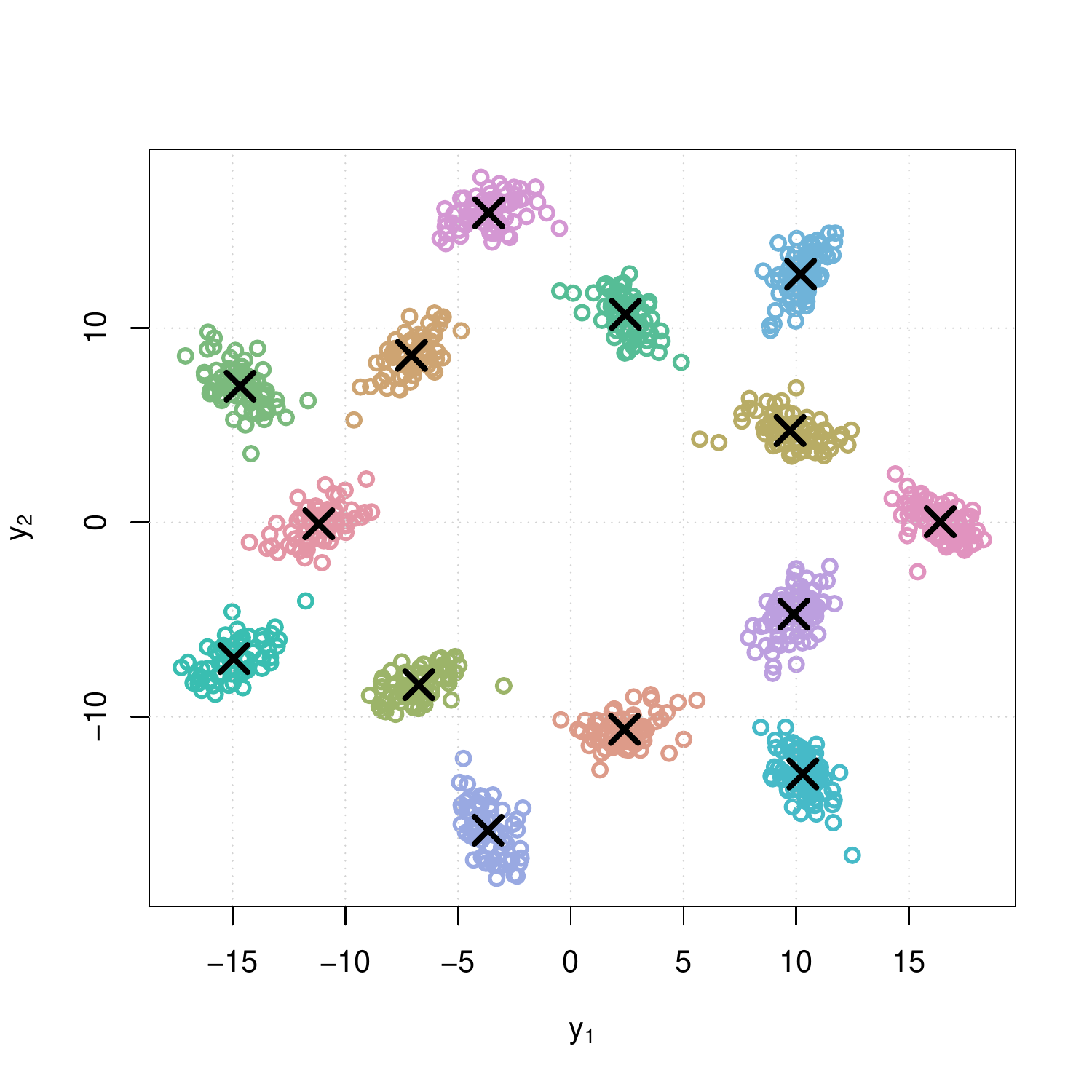}
\par\end{centering}
\caption{\label{fig wreath}Plot of the 1000 observations of the Wreath data
set, colored by subpopulation with subpopulation means indicated by
crosses.}

\end{figure}

The Wreath data (accessed in $\texttt{R}$ via the $\texttt{data(wreath)}$
command) contain 1000 observations of $d=2$ dimensional vectors,
each belonging to one of $g=14$ distinct but unlabelled subpopulations.
We use the $\texttt{Mclust}$ function from $\texttt{mclust}$ to
fit a 14 component mixture normal distributions to the data. The data,
along with the means of the subpopulation normal distributions, are
plotted in Figure \ref{fig wreath}. Here, each observation is colored
based upon the subpopulation that maximizes its \emph{a posteriori}
probability.

As with the Iris data, using the fitted mixture model as a template,
we can then simulate synthetic data sets of any size $n$. We perform
two experiments using this scheme. In the first experiment, we simulate
$n=10^{6}$ observations and assess the different algorithms, based
on the computation time, the log-likelihood, the SE, and the ARI over
$\text{Rep=100}$ repetitions, as per Iris1. We refer to this experiment
as Wreath1. In the second experiment, we repeat the setup of Wreath1,
but with $n=10^{7}$, instead. We refer to this case as Wreath2.

\subsection{Results}

\begin{figure}
\begin{centering}
\subfloat[Timing results, in seconds.]{\begin{centering}
\includegraphics[width=10cm]{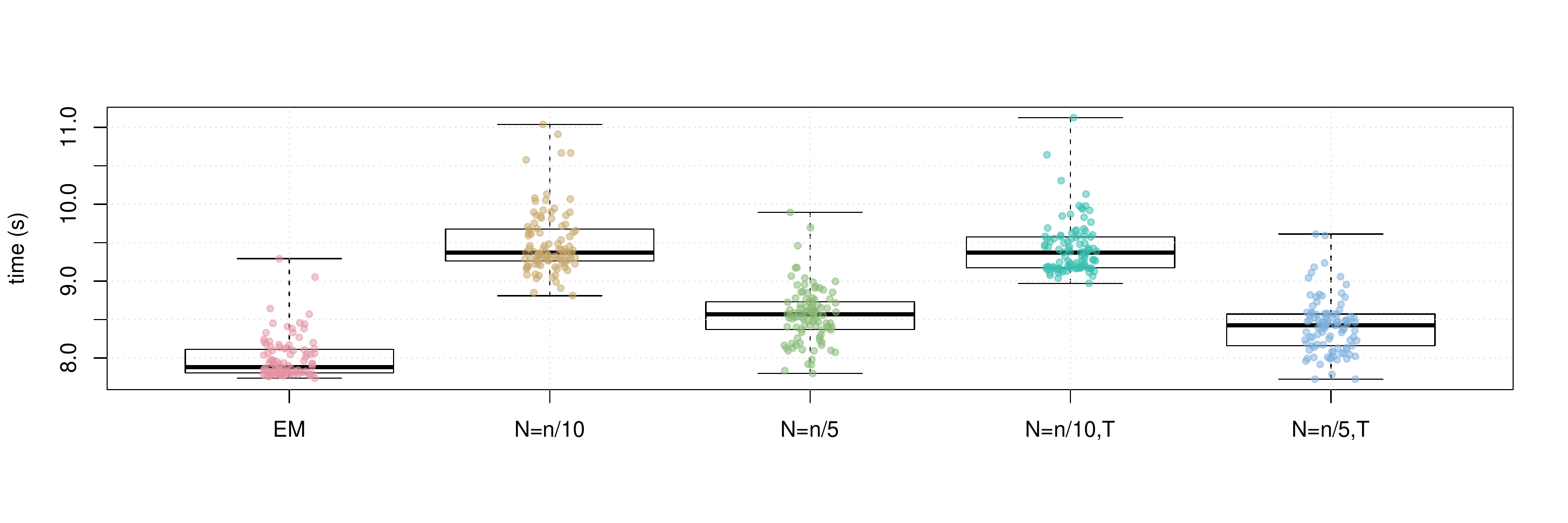}
\par\end{centering}
\centering{}}
\par\end{centering}
\begin{centering}
\subfloat[Log-likelihood results.]{\begin{centering}
\includegraphics[width=10cm]{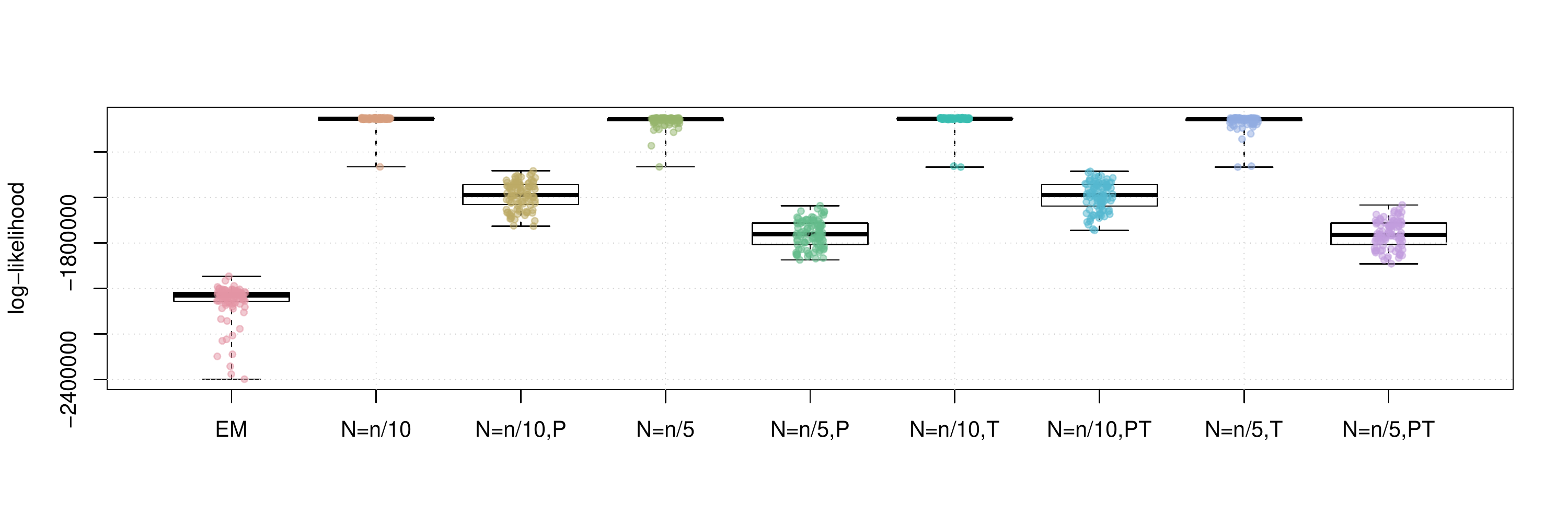}
\par\end{centering}
\centering{}}
\par\end{centering}
\begin{centering}
\subfloat[Standard error results.]{\begin{centering}
\includegraphics[width=10cm]{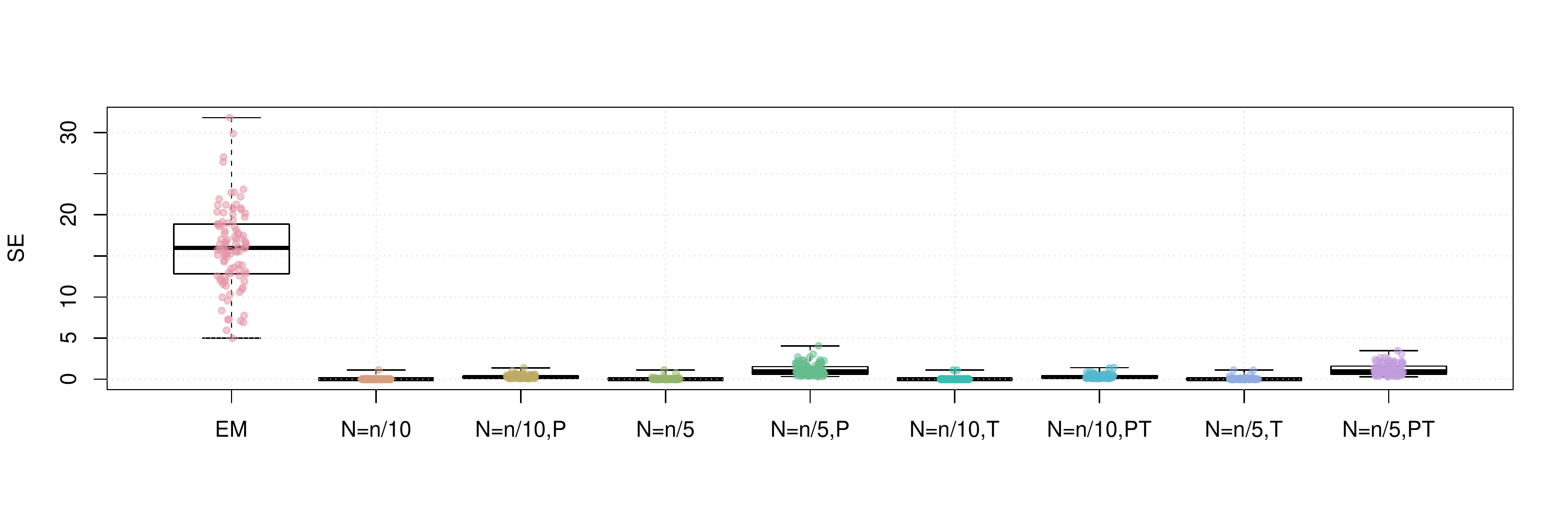}
\par\end{centering}
\centering{}}
\par\end{centering}
\begin{centering}
\subfloat[Adjusted-Rand index results.]{\begin{centering}
\includegraphics[width=10cm]{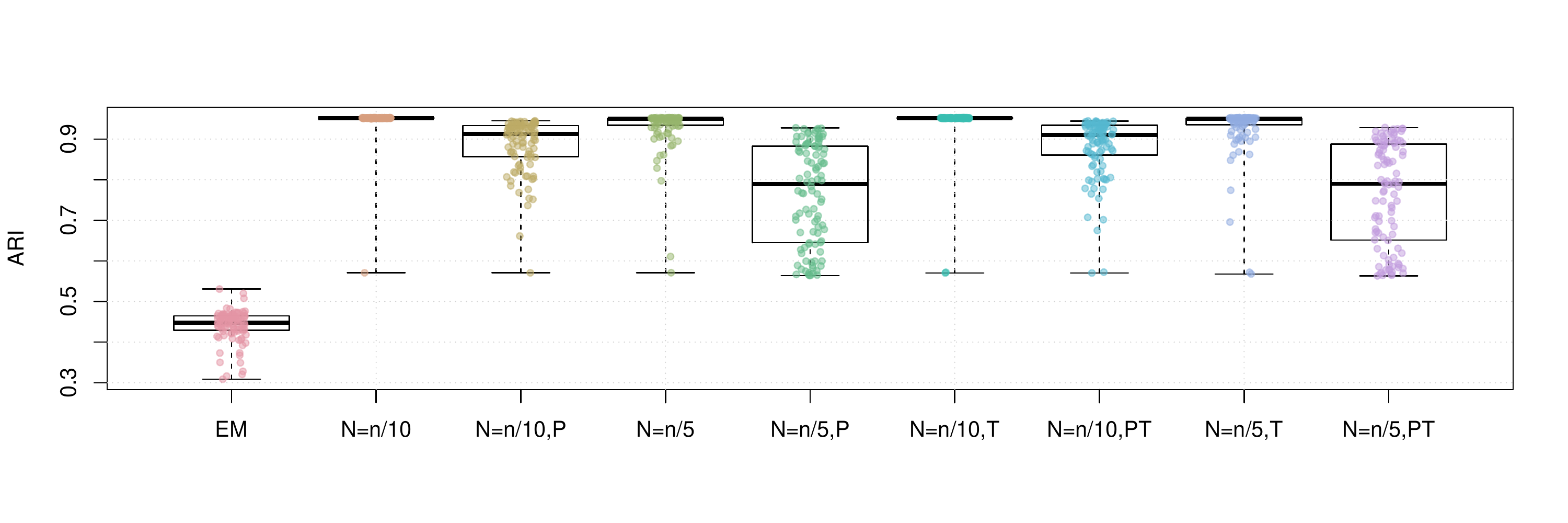}
\par\end{centering}
\centering{}}
\par\end{centering}
\caption{\label{fig: iris1}Results from $\text{Rep}=100$ replications of
the Iris1 simulation experiment. The 'EM' box plot summarizes the
performance of the standard EM algorithm. The other plots are labelled
by which variant of the mini-batch EM algorithm is summarized. The
value of the batch size $N$ is indicated (either $N=n/10$ or $N=n/5$),
and a 'P' or a 'T' designates that Polyak averaging or truncation
was used, respectively. }
\end{figure}

\begin{figure}
\begin{centering}
\subfloat[Timing results, in seconds.]{\begin{centering}
\includegraphics[width=10cm]{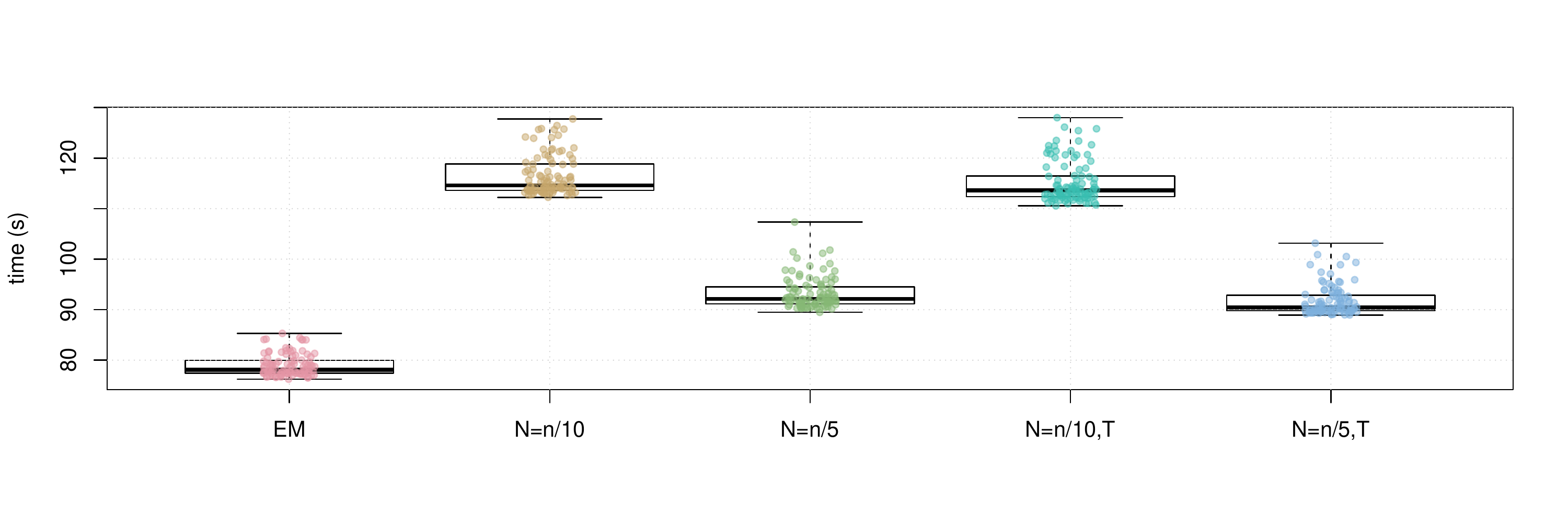}
\par\end{centering}
\centering{}}
\par\end{centering}
\begin{centering}
\subfloat[Log-likelihood results.]{\begin{centering}
\includegraphics[width=10cm]{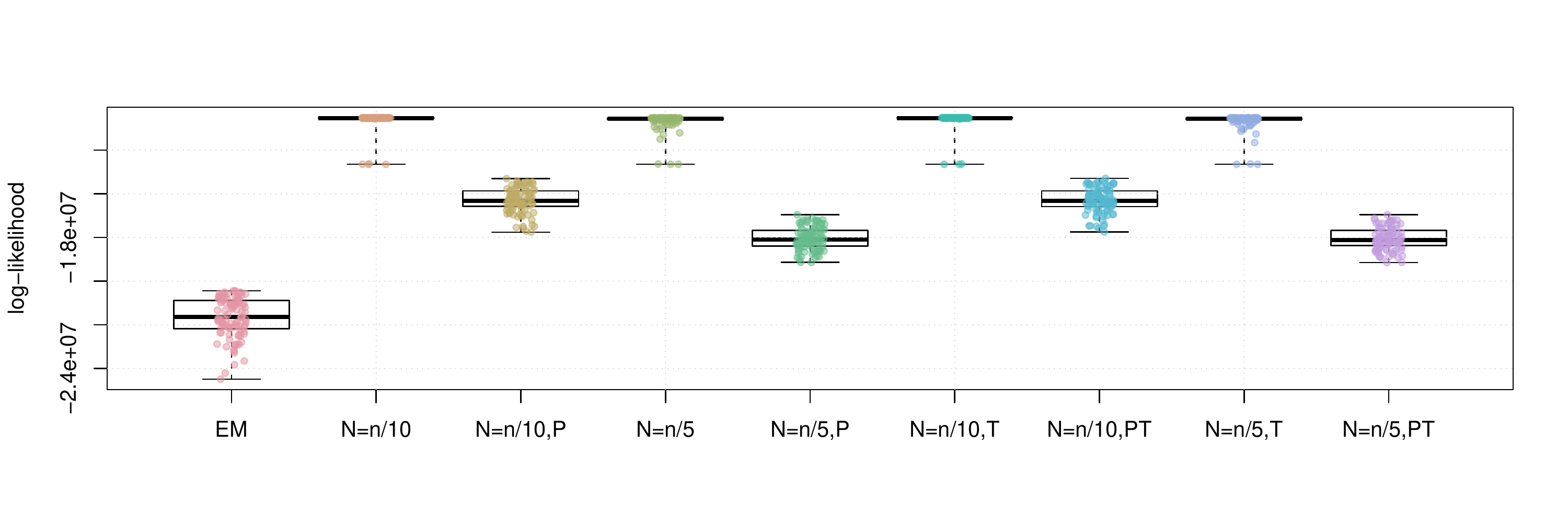}
\par\end{centering}
\centering{}}
\par\end{centering}
\begin{centering}
\subfloat[Standard error results.]{\begin{centering}
\includegraphics[width=10cm]{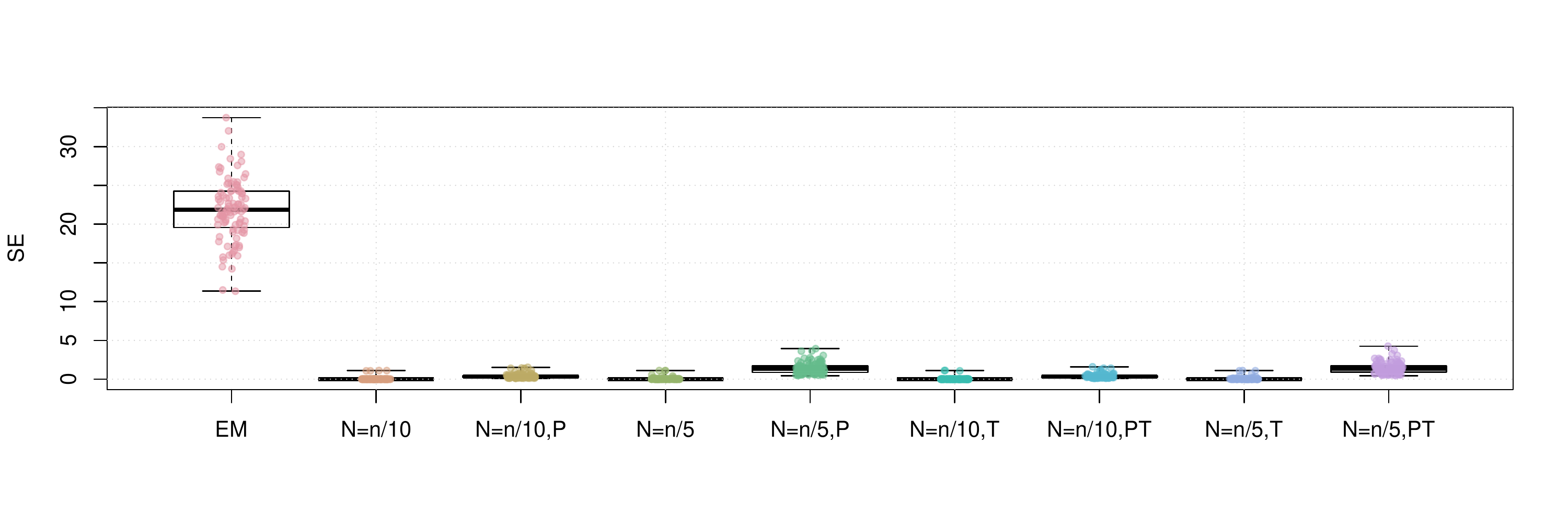}
\par\end{centering}
\centering{}}
\par\end{centering}
\begin{centering}
\subfloat[Adjusted-Rand index results.]{\begin{centering}
\includegraphics[width=10cm]{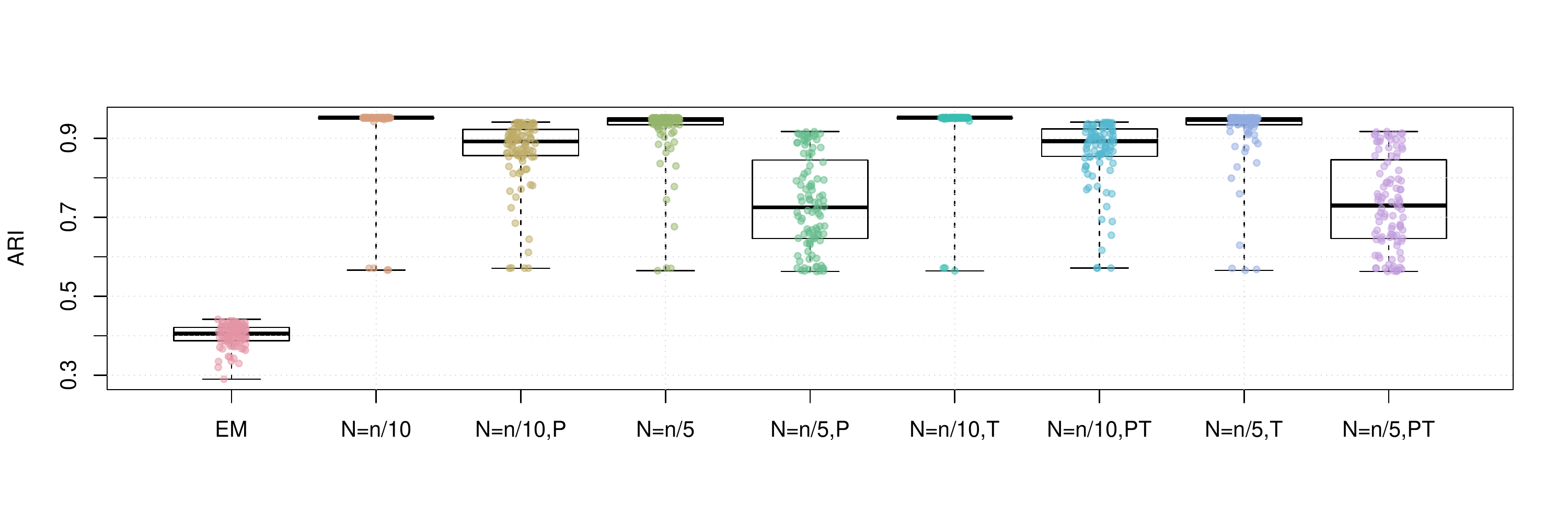}
\par\end{centering}
\centering{}}
\par\end{centering}
\caption{\label{fig: iris2}Results from $\text{Rep}=100$ replications of
the Iris2 simulation experiment. The 'EM' box plot summarizes the
performance of the standard EM algorithm. The other plots are labelled
by which variant of the mini-batch EM algorithm is summarized. The
value of the batch size $N$ is indicated (either $N=n/10$ or $N=n/5$),
and a 'P' or a 'T' designates that Polyak averaging or truncation
was used, respectively.}
\end{figure}

\begin{figure}
\begin{centering}
\subfloat[Timing results, in seconds.]{\begin{centering}
\includegraphics[width=10cm]{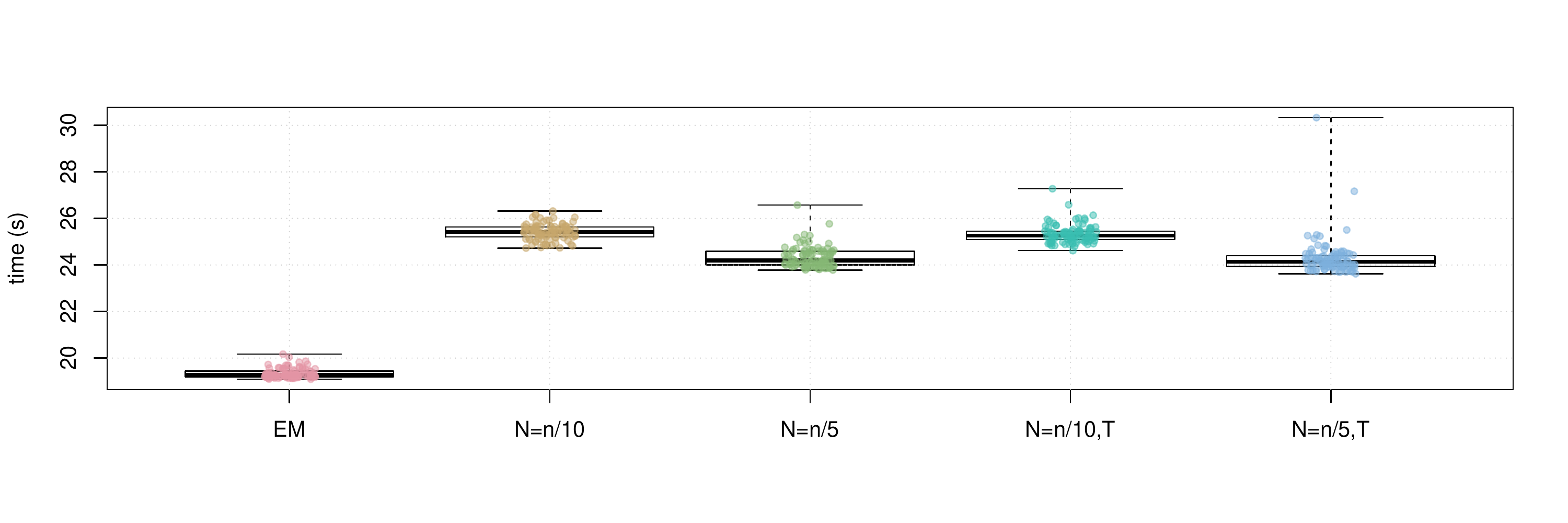}
\par\end{centering}
\centering{}}
\par\end{centering}
\begin{centering}
\subfloat[Log-likelihood results.]{\begin{centering}
\includegraphics[width=10cm]{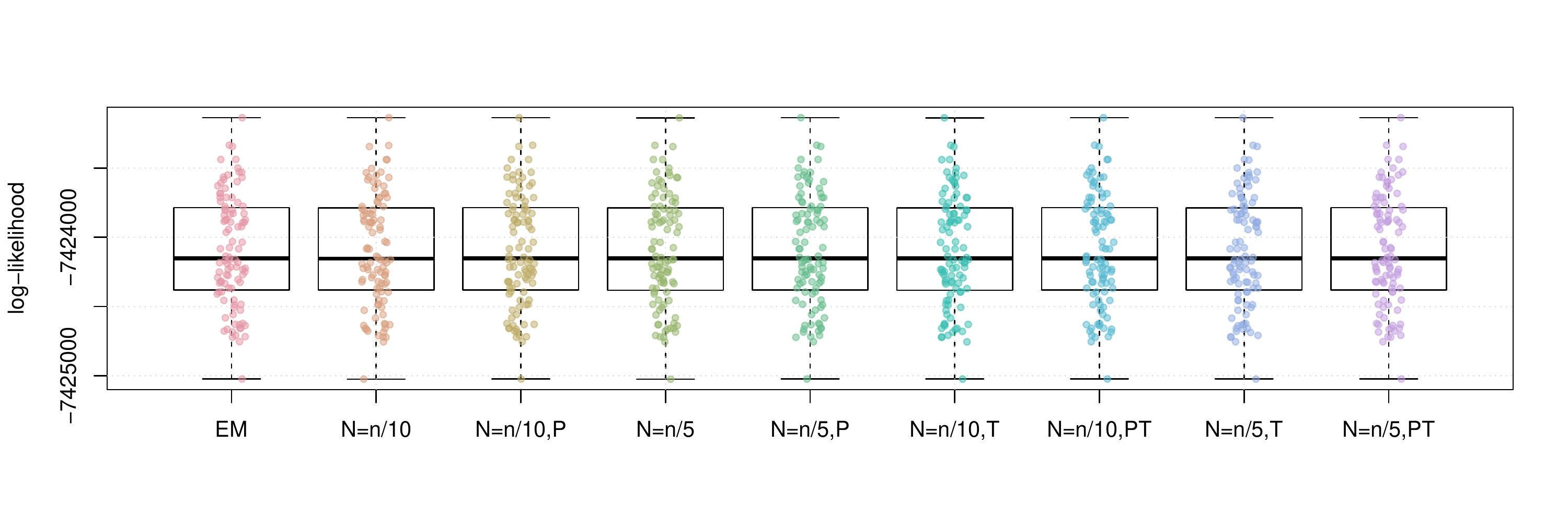}
\par\end{centering}
\centering{}}
\par\end{centering}
\begin{centering}
\subfloat[Standard error results.]{\begin{centering}
\includegraphics[width=10cm]{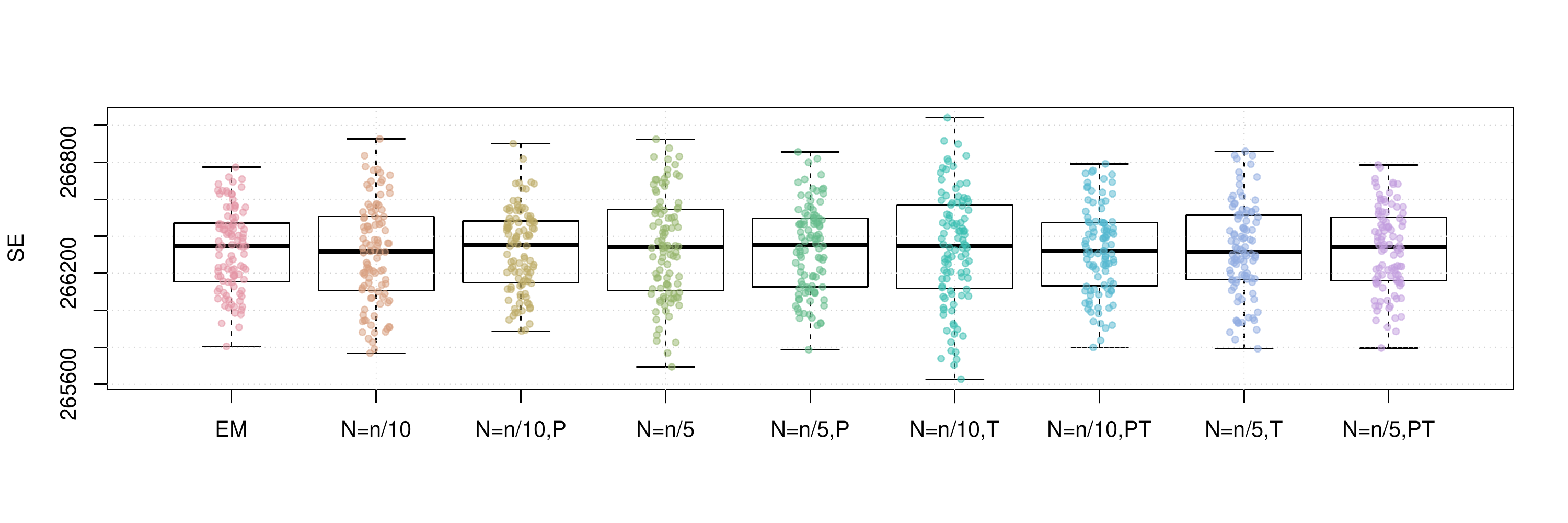}
\par\end{centering}
\centering{}}
\par\end{centering}
\begin{centering}
\subfloat[Adjusted-Rand index results.]{\begin{centering}
\includegraphics[width=10cm]{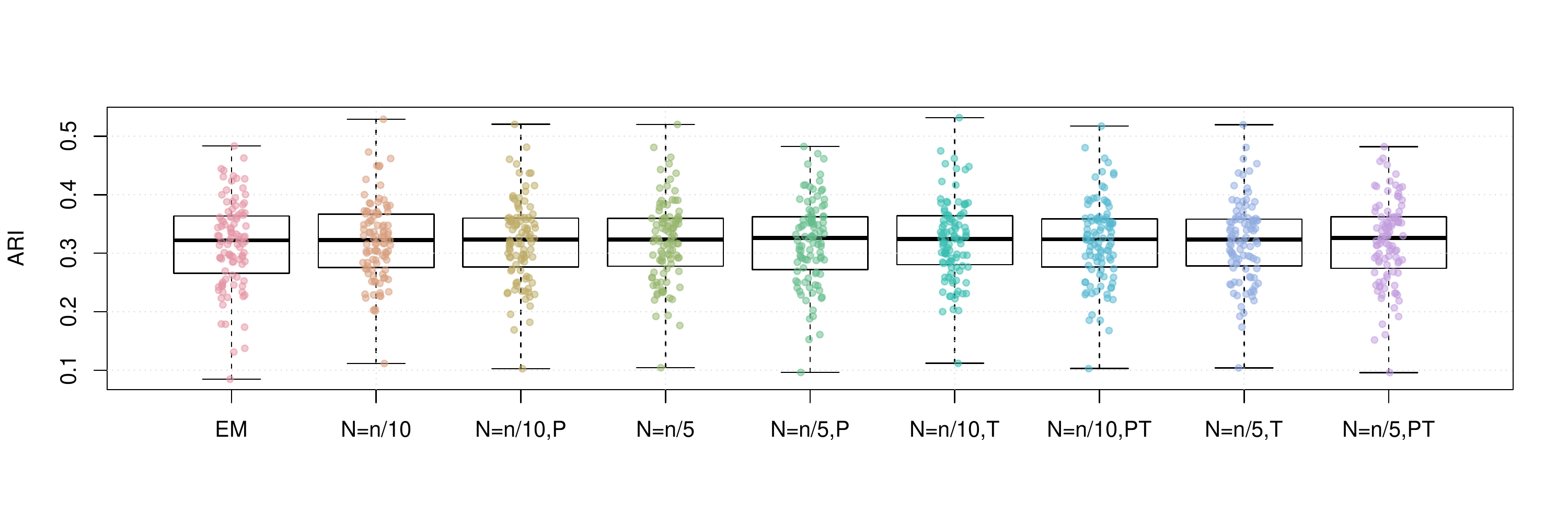}
\par\end{centering}
\centering{}}
\par\end{centering}
\caption{\label{fig: wreath1}Results from $\text{Rep}=100$ replications of
the Wreath1 simulation experiment. The 'EM' box plot summarizes the
performance of the standard EM algorithm. The other plots are labelled
by which variant of the mini-batch EM algorithm is summarized. The
value of the batch size $N$ is indicated (either $N=n/10$ or $N=n/5$),
and a 'P' or a 'T' designates that Polyak averaging or truncation
was used, respectively. }
\end{figure}

\begin{figure}
\begin{centering}
\subfloat[Timing results, in seconds.]{\begin{centering}
\includegraphics[width=10cm]{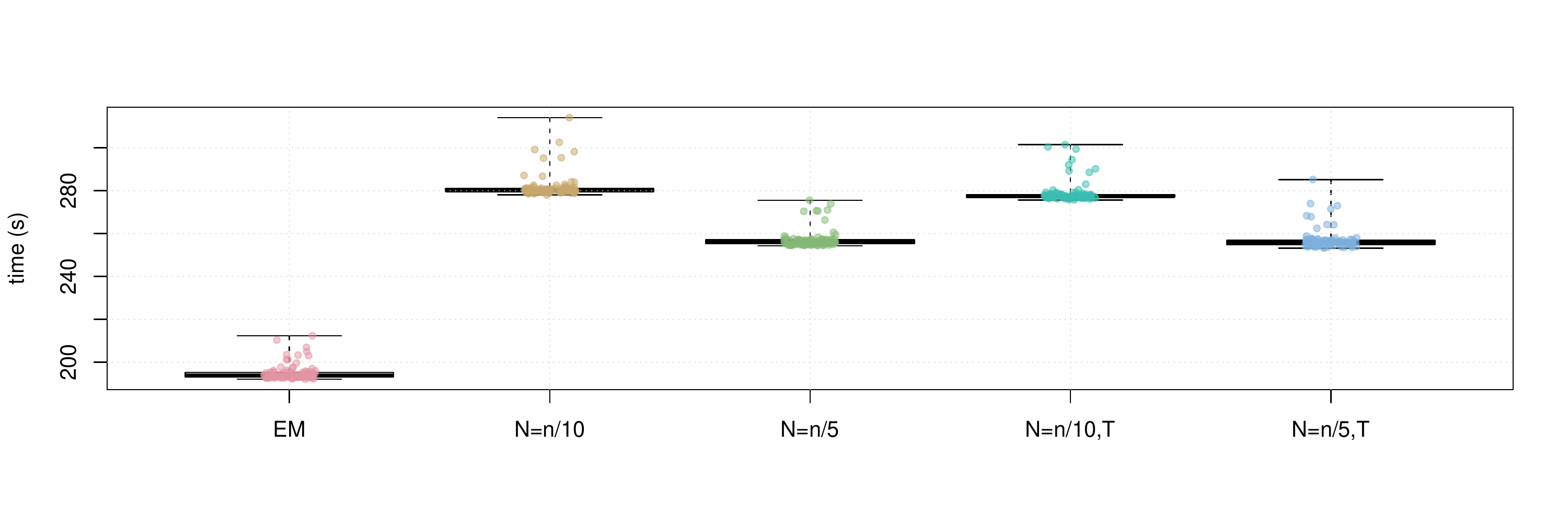}
\par\end{centering}
\centering{}}
\par\end{centering}
\begin{centering}
\subfloat[Log-likelihood results.]{\begin{centering}
\includegraphics[width=10cm]{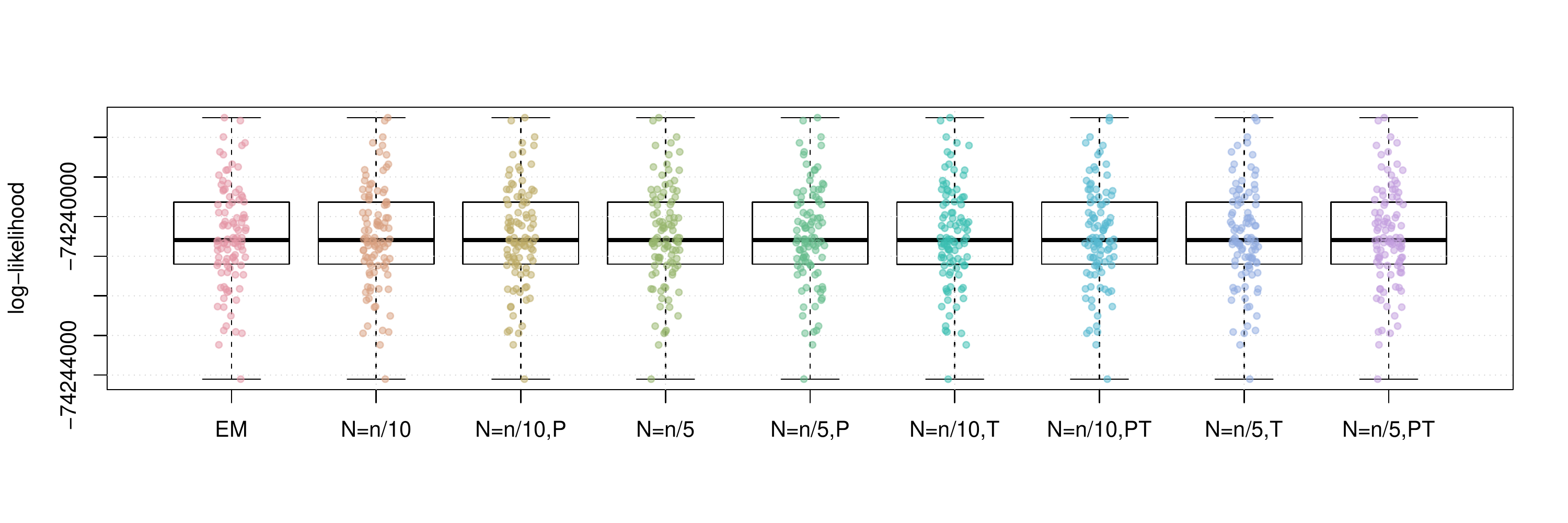}
\par\end{centering}
\centering{}}
\par\end{centering}
\begin{centering}
\subfloat[Standard error results.]{\begin{centering}
\includegraphics[width=10cm]{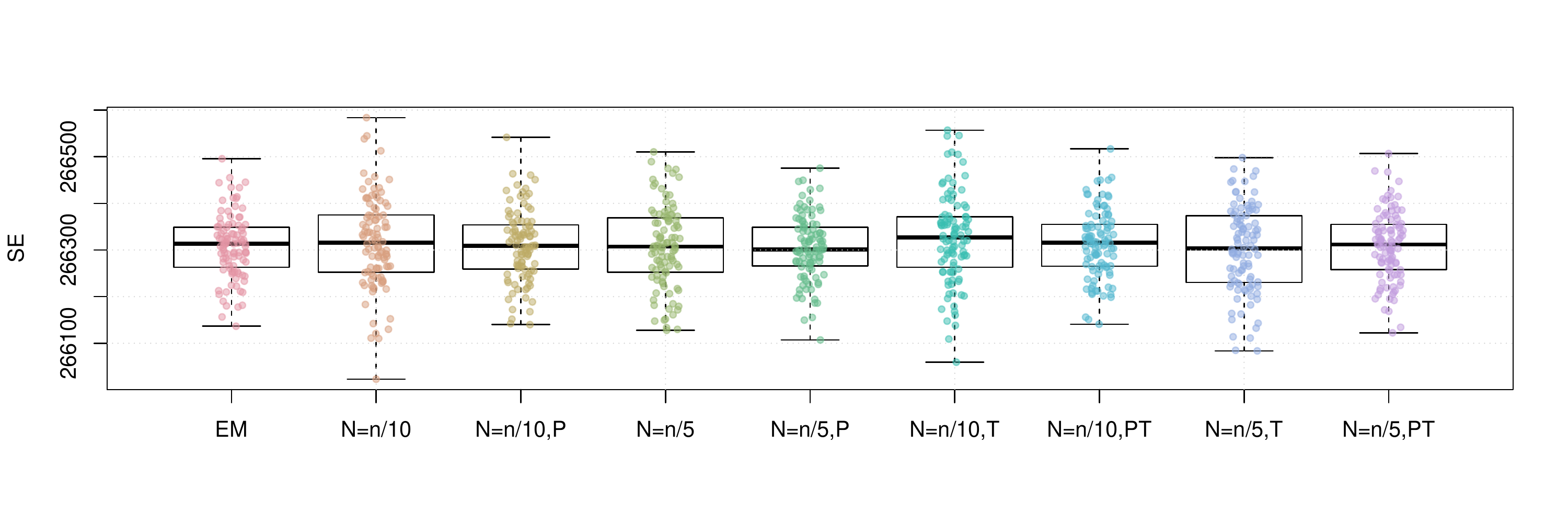}
\par\end{centering}
\centering{}}
\par\end{centering}
\begin{centering}
\subfloat[Adjusted-Rand index results.]{\begin{centering}
\includegraphics[width=10cm]{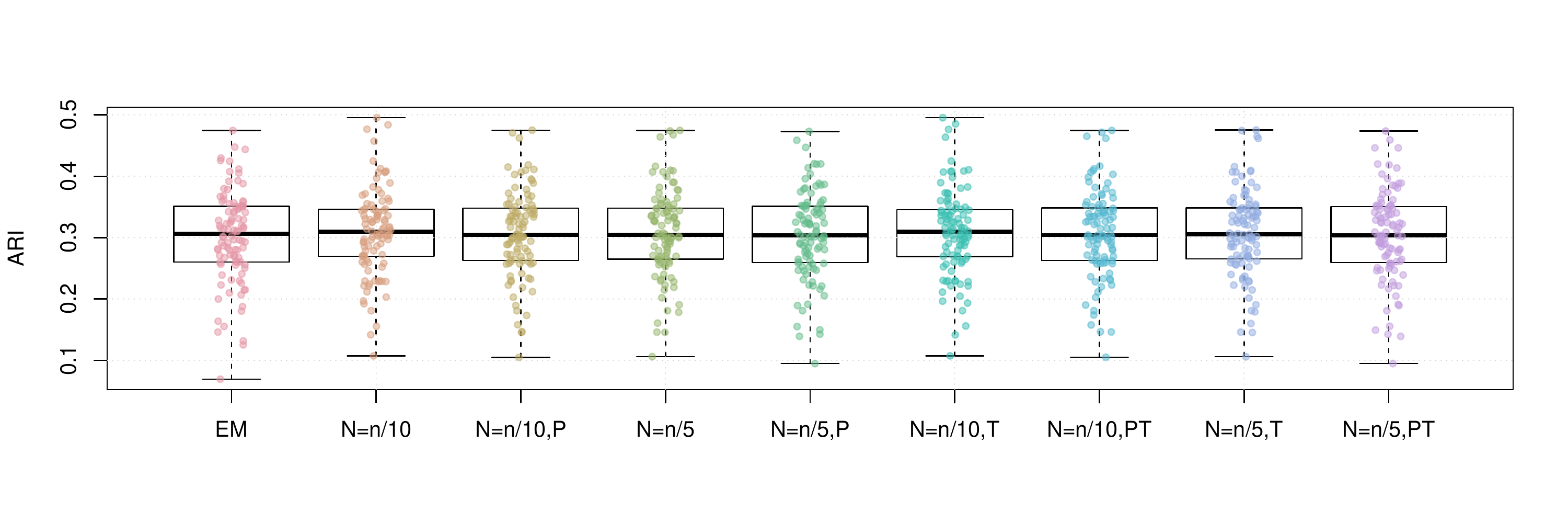}
\par\end{centering}
\centering{}}
\par\end{centering}
\caption{\label{fig: wreath2}Results from $\text{Rep}=100$ replications of
the Wreath2 simulation experiment. The 'EM' box plot summarizes the
performance of the standard EM algorithm. The other plots are labelled
by which variant of the mini-batch EM algorithm is summarized. The
value of the batch size $N$ is indicated (either $N=n/10$ or $N=n/5$),
and a 'P' or a 'T' designates that Polyak averaging or truncation
was used, respectively. }
\end{figure}

Figures \ref{fig: iris1} and \ref{fig: iris2} contain box plots
that summarize the results of Iris1 and Iris2, respectively. Similarly,
Figures \ref{fig: wreath1} and \ref{fig: wreath2} contain box plots
that summarize the results of Wreath1 and Wreath2, respectively.

Firstly, we note that Polyak averaging requires no additional computational
effort, for a given value of $N$. Thus, we do not require separate
timing data for Polyak averaging variants of the mini-batch EM algorithms
in each of Figures \ref{fig: iris1}--\ref{fig: wreath2}. From the
timing results, we observe that the standard EM algorithm is faster
than the mini-batch versions, in all scenarios, regardless of the
fact that all of the algorithms were computed using 10 epochs worth
of data access. This is because the mini-batch algorithms require
more additional intermediate steps in each algorithm loop (e.g., random
sampling from the empirical distribution), as well as a multiplicative
factor of $n/N$ more loops. From the plots, we observe that the larger
value of $N$ tends to result in smaller computing times. There appears
to be no difference in timing between the use of truncation or not.

In the Iris1 and Iris2 studies, we observe that the mini-batch EM
algorithms uniformly outperform the standard EM algorithm in terms
of the log-likelihood, SE, and ARI. In all three measurements, we
observe that $N=n/10$ performed better than $N=n/5$, and also that
there were no differences between truncated versions of the mini-batch
algorithms and equivalent variants without truncation. Polyak averaging
appears to reduce the performance of the mini-batch algorithms, for
a given level of $N$, with respect to each of the three measurements.

In the Wreath1 and Wreath2 studies, we observe that the standard and
mini-batch EM algorithms perform virtually the same across the log-likelihood,
SE, and ARI metrics. This is likely due to the high degree of separability
of each of the $g=12$ mixture components of the Wreath data, in comparison
to the overlapping components of the Iris data. Our observation is
true for all of the mini-batch EM variants, with or without truncation
or Polyak averaging. Upon first impression, this may appear as a weakness
of the mini-batch EM algorithm, since it produces the same performance
while requiring more computational time. However, we must also remember
that the mini-batch EM algorithm does not require all of the data
to be stored in memory at each iteration of algorithm, whereas the
EM algorithm does. Thus, the mini-batch algorithm is feasible in very
large data situations, since it only requires a fixed memory size,
of order $N$, regardless of sample size $n$, whereas the standard
EM algorithm has a memory requirement that increases with $n$.

To expand upon our currently presented results, we have also included
a further two simulation studies regarding the fitting of finite mixtures
of normal distributions using mini-batch EM algorithms. Our two studies
are based on the Flea data of \citet{Wickham:2011aa}, a test scenario
from the ELKI project of \citet{Schubert:2015aa}, and an original
data generating process. The ELKI scenario is chosen due to its separability,
in order to assess whether our conclusion regarding Wreath1 and Wreath2
are correct. The Flea data shares similarities with the Iris data
but is higher dimensional. In all cases, we found that the mini-batch
algorithms tended to outperform the EM algorithm in all but timing,
on average. Detailed assessments of these studies can be found in
the Supplementary Materials.

In addition, we have also investigated the use of the mini-batch EM
algorithm for estimation of non-normal mixture models. Namely, we
present a pair of algorithms for the estimation of exponential and
Poisson mixture models. We demonstrate their performance via an additional
pair of simulation studies.

To conclude, we make the following recommendations. Firstly, smaller
batch sizes appear to yield higher likelihood values. Secondly, averaging
appears to slow convergence of the algorithm to the higher likelihood
value and is thus not recommended. Thirdly, truncation appears to
have no effect on the performance. This is likely due to the fact
that truncation may not have been needed in any of the experiments.
In any case, it is always useful to use the truncated version of the
algorithm, in case there are unforeseen instabilities in the optimization
process. And finally, the standard EM algorithm may be preferred to
the mini-batch EM algorithm when sample sizes are small and when the
data are highly separable. However, even in the face of high separability,
for large $n$, it may not be feasible to conduct estimation by the
standard EM algorithm and thus the mini-batch algorithms may be preferred
due to feasibility.

It is interesting to observe that Polyak averaging tended to diminish
the performance of the algorithms, in our studied scenarios. This
is in contradiction to the theory that suggests that Polyak averaging
should in fact increase the convergence rate to stationary solutions.
We note, however, that the theory is asymptotic and the number of
epochs that were used may be too short for the advantages of Polyak
averaging to manifest, in practice.

\section{\label{sec:Real-data-study}Real data study}

\subsection{MNIST data}

The MNIST data of \citet{LeCun1998} consists of $n=70,000$ observations
of $d=28\times28=784$ pixel images of handwritten digits. These handwritten
digits were sampled nearly uniformly. That is there were 6903, 7877,
6990, 7141, 6824, 6313, 6876, 7293, 6825, and 6958 observations of
the digits 0--9, respectively.

Next, it is notable that not all $d$ pixels are particularly informative.
In fact, there is a great amount of redundancy in the $d$ dimensions.
Out of the $d$ pixels, 65 are always zero, for every observation.
Thus, the dimensions of the data are approximately 8.3\% sparse.

We eliminate the spare pixels across all images to obtain a dense
dimensionality of $d_{\text{dense}}=719$. Using the $d_{\text{dense}}$
dimensions of the data, we conduct a principal component analysis
(PCA) in order to further reduce the data dimensionality; see \citealp{Jolliffe:2002aa}
for a comprehensive treatment on PCA. Using the PCA, we extract the
principal components (PCs) of each observation, and for various number
of PCs $d_{\text{PC}}\in\left[d_{\text{dense}}\right]$. We can then
use the data sets of $n$ observations and dimension $d_{\text{PC}}$,
to estimate mixture of normal distributions for various values of
$g$.

\subsection{Experimental setup}

In the following study, we utilize only the truncated version of the
mini-batch algorithm, having drawn the conclusions, from Section \ref{sec:Simulation-studies},
that there appeared to be no penalty in performance due to truncation
in practice. Again, drawing upon our experience from Section 4, we
set $N=n/10=7000$ as the batch size in all applications. The same
learning rate sequence of $\left\{ \gamma_{r}\right\} _{r=1}^{\infty}$,
where $\gamma_{r}=\left(1-10^{-10}\right)\times r^{6/10}$ is also
used, and $c_{1},c_{2},c_{3}=1000$.

We apply the mini-batch algorithm to data with $d_{\text{PC}}=10,20,50,100$.
Initialization of the parameter vector $\bm{\theta}^{\left(0\right)}$
was conducted via the randomization scheme of \citet[Sec. 3.9.3]{McLachlan2000}.
The mini-batch algorithm was run 100 times for each $d_{\text{PC}}$
and the log-likelihood values were recorded for both the fitted models
using the Polyak averaging and no averaging versions of the algorithm.
The standard EM algorithm, as applied via the $\text{\texttt{em}}$
function of the $\texttt{mclust}$ package is again used for comparison.
Each of the algorithms, including the $k\text{-means}$ algorithm,
were initialized from the same initial randomization, in the interest
of fairness, for each of the 100 runs. That is, a random partition
of the data is generated once for each of the 100 runs, and the initial
parameters for the EM, mini-batch and $k\text{-means}$ algorithms
are all computed from the same initialization. The log-likelihood
values of the standard and mini-batch EM algorithms are compared along
with the ARI values. Algorithms are run for 10 epochs.

We compute the ARI values obtained when comparing the maximum \emph{a
posteriori }clustering labels, obtained from each of the algorithms
(cf. \citealp[Sec. 1.15]{McLachlan2000}), and the true digit classes
of each of the images. For a benchmark, we also compare the performance
of the three EM algorithms with the $k\text{-means}$ algorithm, as
applied via the $\texttt{kmeans}$ function in $\texttt{R}$, which
implements the algorithm of \citet{HartiganWong1979}. For fairness
of comparison, we also allow the $k\text{-means}$ algorithm 10 epochs
in each of 100 runs. As in Section \ref{sec:Simulation-studies},
we note that all codes are available at \url{https://github.com/hiendn/StoEMMIX},
for the sake of reproducibility and transparency.

\subsection{Results}

The results from the MNIST experiment are presented in Table \ref{tab: MNIST}.
We observe that for $d_{\text{PC}}\in\left\{ 10,20,50\right\} $,
all three EM variants provided better ARI than the $k\text{-means}$
algorithm. The best ARI values for all three EM algorithms occur when
$d_{\text{PC}}=20$. When $d_{\text{PC}}=100$, the $k\text{-means}$
algorithm provided a better ARI, which appeared to be somewhat uniform
across the four values of $d_{\text{PC}}$.

Among the EM algorithms, the mini-batch algorithm provided better
ARI values, with the two variants not appearing to be significantly
different from one another, when considering the standard errors of
the ARI values, when $d_{\text{PC}}\in\left\{ 20,50\right\} $. When
$d_{\text{PC}}=10$, we observe that no averaging yielded a better
ARI, whereas, when $d_{\text{PC}}=100$, averaging appeared to be
better, on average.

Regarding the log-likelihoods, the mini-batch EM algorithm, when applied
without averaging, uniformly and significantly outperformed the standard
EM algorithm. On the contrary, when applied with averaging, the EM
algorithm uniformly and significantly outperformed the mini-batch
algorithm. This is also in contrary with what was observed in Section
\ref{sec:Simulation-studies}. This is an interesting result considering
that the ARI of the mini-batch algorithm, with averaging, is still
better than that of the EM algorithm. As in Section \ref{sec:Simulation-studies},
we can recommend the use of the mini-batch EM algorithm without averaging,
as it tends to outperform the standard EM algorithm for fit and is
also yields better clustering outcomes, when measured via the ARI.

\begin{table}
\caption{\label{tab: MNIST}Tabulation of results from the 100 runs of the
EM algorithms and the $k\text{-means}$ algorithm, for each value
of $d_{\text{PC}}\in\left\{ 10,20,50,100\right\} $. The columns EM,
Mini, and Mini Pol refer to the standard EM, the mini-batch EM, and
the mini-batch EM algorithm with Polyak averaging, respectively. The
SE rows contain the standard error over each of the 100 runs (i.e.
the standard deviation over 10). Boldface text highlight the best
results.}

\centering{}%
\begin{tabular}{|rr|rrrr|rrr|}
\hline 
 &  & ARI &  &  &  & log-likelihood &  & \tabularnewline
$d_{\text{PC}}$ &  & EM & Mini & Mini Pol & $k\text{-means}$ & EM & Mini & Mini Pol\tabularnewline
\hline 
10 & Mean & 0.401 & \textbf{0.443} & 0.432 & 0.352 & -4.98E+06 & \textbf{-4.96E+06} & -5.01E+06\tabularnewline
 & SE & 0.004 & 0.004 & 0.004 & 0.002 & 1.19E+03 & 6.43E+02 & 5.90E+02\tabularnewline
\hline 
20 & Mean & 0.436 & 0.475 & \textbf{0.480} & 0.367 & -9.46E+06 & \textbf{-9.44E+06} & -9.52E+06\tabularnewline
 & SE & 0.005 & 0.005 & 0.005 & 0.002 & 2.20E+03 & 1.37E+03 & 1.67E+03\tabularnewline
\hline 
50 & Mean & 0.394 & 0.434 & \textbf{0.438} & 0.369 & -2.18E+07 & \textbf{-2.17E+07} & -2.20E+07\tabularnewline
 & SE & 0.005 & 0.005 & 0.006 & 0.002 & 8.47E+03 & 7.01E+03 & 4.85E+03\tabularnewline
\hline 
100 & Mean & 0.326 & 0.356 & \textbf{0.377} & 0.372 & -3.99E+07 & \textbf{-3.97E+07} & -4.05E+07\tabularnewline
 & SE & 0.004 & 0.004 & 0.005 & 0.002 & 1.83E+04 & 1.68E+04 & 8.72E+03\tabularnewline
\hline 
\end{tabular}
\end{table}

\section{Conclusions}

In Section \ref{sec:The-minibatch-EM}, we reviewed the online EM
algorithm framework of \citet{Cappe2009}, and stated the key theorems
that guarantee the convergence of algorithms that are constructed
under the online EM framework. We then presented a novel interpretation
of the online EM algorithm that yielded our framework for constructing
mini-batch EM algorithms. We then utilized the theorems of \citet{Cappe2009}
in order to produce convergence results for this new mini-batch EM
algorithm framework. Extending upon some remarks of \citet{Cappe2009},
we also made rigorous the use of truncation in combination with both
the online EM and mini-batch EM algorithm frameworks, using the construction
and theory of \citet{Delyon1999}.

In Section \ref{sec:Normal-mixture-models}, we demonstrated how the
mini-batch EM algorithm framework could be applied to construct algorithms
for conducting ML estimation of finite mixtures of exponential family
distributions. A specific analysis is made of the particularly interesting
case of the normal mixture models. Here, we validate the conditions
that permit the use of the Theorems from Section \ref{sec:The-minibatch-EM}
in order to guarantee the convergence of the mini-batch EM algorithms
for ML estimation of normal mixture models.

In Section \ref{sec:Simulation-studies}, we conducted a set of four
simulation studies in order to study the performance of the mini-batch
EM algorithms, implemented in eight different variants, as compared
to the standard EM algorithm for ML estimation of normal mixture models.
There, we found that regardless of implementation, in many cases,
the mini-batch EM algorithms were able to obtain log-likelihood values
that were better on average than the standard EM algorithm. We also
found that the use of larger batch sizes and Polyak averaging tended
to diminish performance of the mini-batch algorithms, but the use
of truncation tended to have no effect. Although the mini-batch algorithms
is generally slower than the standard EM algorithm, we note that in
many cases, the fixed memory requirement of the mini-batch algorithms
make them feasible where the standard EM algorithm is not.

A real data study was conducted in Section \ref{sec:Real-data-study}.
There, we explored the use of the standard EM algorithm and the truncated
mini-batch EM algorithm for cluster analysis of the famous MNIST data
of \citet{LeCun1998}. From our study, we found that the mini-batch
EM algorithm was able to obtain better log-likelihood values than
the standard EM algorithm, when applied without Polyak averaging.
However, with averaging, the mini-batch EM algorithm was worse than
the standard EM algorithm, on average. However, regardless of whether
averaging was used, or not, the mini-batch EM algorithm appeared to
yield better clustering outcomes, when measured via the ARI of \citet{Hubert1985}.

This research poses numerous interesting directions for the future.
First, we may extend the results to other exponential family distributions
that permit the satisfaction of theorem assumptions from Section \ref{sec:The-minibatch-EM}.
We make initial steps in this direction via a pair of mini-batch algorithms
for exponential and Poisson distribution mixtures. Secondly, we may
use the framework to construct mini-batch algorithms for large-scale
mixture of regression models (cf. \citealp{Jones1992}), following
the arguments made by \citet{Cappe2009} that permitted them to construct
an online EM algorithm for their mixture of regressions example analysis.
Thirdly, this research theme can be extended further to the construction
of mini-batch algorithms for mixture of experts models (cf. \citealp{Nguyen2018}),
which may be facilitated via the Gaussian gating construction of \citet{Xu1995}.

In addition to the three previous research questions, we may also
ask questions regarding the practical application of the mini-batch
algorithms. For instance, we may consider the question of optimizing
learning rates and batch sizes for particular application settings.
Furthermore, we may consider whether the theoretical framework still
applies to algorithms where we may have adaptive batch sizes and learning
rate regimes. As these directions fall vastly outside the scope of
the current paper, we shall leave them for future exploration.

\section*{Acknowledgements}

The authors are indebted to the Co-ordinating Editor and two Reviewers
for their insightful comments that have improved the exposition of
the manuscript. HDN is personally funded by Australian Research Council
(ARC) grant DE170101134. GJM and HDN are also funded under ARC grant
DP180101192. The work is supported by Inria project LANDER.

\bibliographystyle{apalike2}
\bibliography{MASTERBIB}

\end{document}